\documentclass[prd,preprint,superscriptaddress,preprintnumbers,eqsecnum,showpacs,nofootinbib,
nobibnotes,noeprint]{revtex4-1}
\usepackage{amsfonts,bm}
\usepackage{amsfonts,amssymb,amsmath}
\usepackage{graphicx}
\usepackage{paralist}
\usepackage[english]{babel}
\usepackage{slashed}
\usepackage[usenames]{xcolor}
\usepackage{mathtools}
\usepackage[makeroom]{cancel}
\usepackage{color}

\usepackage[linkcolor=blue,citecolor=blue,urlcolor=blue,colorlinks=true]{hyperref}

\newcommand{\be}{\begin{equation}}
\newcommand{\bea}{\begin{eqnarray}}
\newcommand{\ee}{\end{equation}}
\newcommand{\eea}{\end{eqnarray}}

\def\1eq#1{Eq.~(\ref{#1})}

\def\2eqs#1#2{Eqs.~(\ref{#1}) and~(\ref{#2})}
\def\3eqs#1#2#3{Eqs.~(\ref{#1}),~(\ref{#2}) and~(\ref{#3})}

\def\fig#1{Fig.~\ref{#1}}

\def\ie{{\it i.e.}, }
\def\eg{{\it e.g.}, }

\newcommand{\Ls}{ \mathit{L}_{{sg}}}   
\def\g{\Gamma}
\def\gz{\Gamma_{\!0}}

\def\s#1{{\scriptscriptstyle #1}}

\newcommand{\fatg}{{\rm{I}}\!\Gamma}
\newcommand{\fatgt}{\widetilde{{\rm{I}}\!\Gamma}}

\newcommand{\w}{{\cal W}}
\newcommand{\kw}{{K}}

\newcommand{\Ln}{\mathit{L}_0}
\newcommand{\wn}{{\w}_0}

\newcommand{\Cfat}{{\mathbb C}}
\newcommand{\Cfattilde}{\widetilde{\mathbb C}}
\newcommand{\C}{{\mathcal C}}
\newcommand{\Ctilde}{{\widetilde{\mathcal C}}}

\newcommand{\wB}{{\w}_{\star}}
\newcommand{\CB}{{\mathbb C}_{\star}}
\newcommand{\Cc}{{\mathcal C}_{\star}}

\newcommand{\Cgh}{C}

\newcommand{\ws}{{\w}_{{sg}}}

\newcommand{\bL}{ \bar{\mathit{L}}_{{sg}}}
\newcommand{\Ical}{ {\cal I} }

\begin{document}

\title{Exploring smoking-gun signals\\ of the Schwinger mechanism in QCD}

\author{A.~C. Aguilar}
\affiliation{\mbox{University of Campinas - UNICAMP, Institute of Physics ``Gleb Wataghin'',} \\
13083-859 Campinas, S\~{a}o Paulo, Brazil}

\author{M.~N. Ferreira}
\affiliation{\mbox{University of Campinas - UNICAMP, Institute of Physics ``Gleb Wataghin'',} \\
13083-859 Campinas, S\~{a}o Paulo, Brazil}

\author{J. Papavassiliou}
\affiliation{\mbox{Department of Theoretical Physics and IFIC, 
University of Valencia and CSIC},
E-46100, Valencia, Spain}

\begin{abstract}

In QCD, the Schwinger mechanism endows the gluons with an effective mass  
through the dynamical formation of massless bound-state poles that are longitudinally coupled.
The presence of these poles  
affects profoundly the infrared properties of the interaction vertices,  
inducing crucial modifications to their fundamental Ward identities.
Within this general framework,  
we present a detailed derivation of the non-Abelian Ward identity obeyed by the pole-free part of the three-gluon vertex
in the soft-gluon limit, and determine the  
smoking-gun displacement that the onset of the Schwinger mechanism produces to the standard result. 
Quite importantly, the quantity that describes this distinctive feature coincides formally
with the bound-state wave function that controls the massless pole formation. Consequently, this
signal may be computed in two independent ways: by solving an approximate version of  
the pertinent Bethe-Salpeter integral equation, or by appropriately combining the
elements that enter in the aforementioned  Ward identity. 
For the implementation of both methods we employ 
two- and three-point correlation functions  obtained from recent lattice simulations, 
and a partial derivative of the ghost-gluon kernel, which is computed from the corresponding 
Schwinger-Dyson equation. Our analysis reveals an excellent coincidence
between the results obtained through either method, providing a highly nontrivial self-consistency check
for the entire approach. When compared to the null hypothesis, where the Schwinger mechanism is assumed to
be inactive, the statistical significance of the resulting signal is estimated to be three standard deviations.

\end{abstract}

\pacs{
12.38.Aw,  
12.38.Lg, 
14.70.Dj 
}

\maketitle

\section{\label{sec:intro}Introduction}

The systematic study of the fundamental $n$-point  correlation (Green's) functions,
such as propagators and vertices,  forms an essential element in the
ongoing quest for unraveling the nonperturbative properties
and underlying dynamical mechanisms of QCD~\cite{Marciano:1977su}.
In recent years, this challenging problem
has been tackled by means of approaches formulated in the continuum,   
such as the  Schwinger-Dyson equations (SDEs)~\cite{Roberts:1994dr,Alkofer:2000wg,Fischer:2006ub,Roberts:2007ji,Binosi:2009qm,Binosi:2014aea,Cloet:2013jya,Aguilar:2015bud,Binosi:2016rxz,Binosi:2016nme,Huber:2018ned}  or the functional renormalization group~\mbox{\cite{Pawlowski:2003hq,  Pawlowski:2005xe,Cyrol:2017ewj, Corell:2018yil, Blaizot:2021ikl}},
in conjunction with numerous gauge-fixed lattice simulations~\mbox{\cite{Cucchieri:2007md,Cucchieri:2007rg,Cucchieri:2009zt,Bogolubsky:2007ud,Bogolubsky:2009dc,Oliveira:2009eh,Oliveira:2010xc,Maas:2011se,Boucaud:2011ug,Oliveira:2012eh}}.
This intense activity has delivered new   
insights  on  the nature and phenomenology  of the
strong  interactions and has broadened our basic understanding  
of non-Abelian gauge theories~\mbox{\cite{Maris:1997tm, Maris:2003vk, Braun:2007bx,Eichmann:2008ef, Cloet:2008re,Boucaud:2008ky,Eichmann:2009qa,Fister:2013bh,Meyer:2015eta,Eichmann:2016yit,Sanchis-Alepuz:2017mir,Alkofer:2018guy,Gao:2017uox,Souza:2019ylx,Xu:2019sns,Aguilar:2019teb,Huber:2020ngt,Roberts:2020udq,Roberts:2020hiw,Horak:2021pfr,Roberts:2021xnz}}.

In this context, the characteristic feature of infrared saturation displayed by the gluon propagator 
has attracted particular attention, being often
linked with the emergence of a mass gap in the gauge sector of QCD~\cite{Smit:1974je,Cornwall:1981zr,Bernard:1981pg,Bernard:1982my,Donoghue:1983fy,Mandula:1987rh,
 Cornwall:1989gv,Wilson:1994fk,Philipsen:2001ip,Aguilar:2002tc,Aguilar:2004sw,Aguilar:2006gr,Aguilar:2008xm, Fischer:2008uz,Tissier:2010ts,Binosi:2012sj,Serreau:2012cg,Pelaez:2014mxa,Aguilar:2016vin}.
This property has been explored both in large-volume simulations~\cite{Cucchieri:2007md,Cucchieri:2007rg,Bogolubsky:2007ud,Bogolubsky:2009dc,Oliveira:2009eh,Oliveira:2010xc,Cucchieri:2009zt},  
and in various functional approaches~\cite{Aguilar:2008xm,Fischer:2008uz,Binosi:2012sj,Serreau:2012cg,Tissier:2010ts, Aguilar:2016vin, Pelaez:2014mxa,Dudal:2008sp,RodriguezQuintero:2010wy,Pennington:2011xs,Meyers:2014iwa,Siringo:2015wtx,Cyrol:2018xeq}, and is rather general, manifesting itself
in the Landau gauge,  away from it~\cite{Cucchieri:2009kk,Cucchieri:2011pp,Bicudo:2015rma,Epple:2007ut,Campagnari:2010wc,Aguilar:2016ock,Glazek:2017rwe},  and in the presence of dynamical quarks~\cite{Bowman:2007du,Kamleh:2007ud,Ayala:2012pb,Aguilar:2013hoa,Aguilar:2019uob}. 
In general terms, the scalar form factor, $\Delta(q^2)$, of the gluon propagator
reaches a finite nonvanishing value in the deep infrared, and the gluon mass, $m$, 
is identified as $\Delta^{-1}(0) = m^2$.

One of the nonperturbative mechanisms put forth in order to explain this special 
behavior of the gluon propagator is based on a non-Abelian
extension of the well-known Schwinger mechanism~\cite{Schwinger:1962tn,Schwinger:1962tp}.
According to the fundamental observation underlying this
mechanism, if the self-energy  
develops a pole at zero momentum transfer ($q^2 = 0$), then the corresponding
vector meson (gluon) acquires a mass, even if the gauge symmetry forbids
a mass term at the level of the fundamental Lagrangian~\cite{Schwinger:1962tn,Schwinger:1962tp,Jackiw:1973tr,Eichten:1974et}. 

The precise implementation of this idea at the level of the SDE describing the momentum
evolution of $\Delta(q^2)$ requires the inclusion of {\it longitudinally coupled massless poles} 
at the level of the fundamental interaction vertices of the theory~\cite{Aguilar:2011xe,Ibanez:2012zk,Binosi:2017rwj,Aguilar:2017dco,Binosi:2017rwj,Eichmann:2021zuv}.
These poles are produced as massless bound state excitations,
whose formation is governed by a special set of Bethe-Salpeter equations (BSEs)~\cite{Aguilar:2011xe,Ibanez:2012zk,Aguilar:2017dco,Binosi:2017rwj}.
In addition, their presence is crucial for maintaining intact the form of the Slavnov-Taylor identities (STIs)~\cite{Taylor:1971ff,Slavnov:1972fg}
satisfied by the corresponding vertices.
Since the fully dressed vertices enter in the diagrammatic expansion of the
gluon SDE, their massless poles   
end up triggering the Schwinger mechanism, enabling a completely dynamical generation of an effective gluon mass~\mbox{\cite{Aguilar:2011xe,Binosi:2012sj,Ibanez:2012zk,Aguilar:2017dco,Binosi:2017rwj}.} 

It is clearly important to further scrutinize the dynamical picture described above, and identify 
certain characteristic properties 
that would corroborate its validity and discriminate it from alternative dynamical scenarios.
In the present work we explore    
a distinctive signal of the non-Abelian Schwinger mechanism, which is intimately connected
with the three-gluon vertex, and 
has the advantage of being reliably calculable by means of well-established inputs, such as
two- and three-point correlation functions  obtained from large-volume lattice simulations.

The pivotal ideas underlying this study may be summarized as follows.
The massless poles are longitudinally coupled, and therefore 
drop out from ``on-shell'' observables~\mbox{\cite{Jackiw:1973tr,Cornwall:1973ts,Eichten:1974et,Poggio:1974qs,Smit:1974je}}, or from the {\it transversely projected} vertices  
employed in lattice simulations ~\cite{Cucchieri:2006tf,Cucchieri:2008qm,Athenodorou:2016oyh,Duarte:2016ieu,Boucaud:2018xup,Aguilar:2019uob},  where only the {\it pole-free} part of the
corresponding vertex survives. Nonetheless, the imprint
of the poles is invariably encoded into the pole-free part, as may be seen 
by considering the  Ward identity (WI) that this latter part satisfies,
namely the limit of the STI as the gluon momentum in the channel of the pole is taken to zero\footnote{The  standard Takahashi identity of QED, \mbox{$q^{\mu}\Gamma_{\mu}(q,p,p+q) = S^{-1}(p+q) - S^{-1}(p)$},
is an Abelian STI; the corresponding WI, $\Gamma_{\mu}(0,p,-p) = \partial S^{-1}(p)/\partial p^{\mu}$,
is obtained from it by expanding around $q=0$.}.

Since the poles contribute nontrivially to the STIs, the corresponding WI 
involves the standard building blocks (\eg propagators) 
{\it and} a residual contribution with a nontrivial momentum dependence,  
which is directly related to the Schwinger mechanism.
As a result, in that kinematic limit, the relevant
form factor of the pole-free part of the vertex is displaced with respect to the case where the Schwinger mechanism is absent.

The above considerations become particularly relevant in the case of the three-gluon vertex, because
the form factor of its pole-free part has been evaluated rather accurately in recent 
lattice simulations~\cite{Boucaud:2002fx,Boucaud:2003xi,Aguilar:2021lke}.  As a result, the displacement originating from the onset of the Schwinger mechanism, to be denoted by $\Cfat(r^2)$, may be calculated by appropriately combining 
this form factor with all other constituents that enter into the WI of the three-gluon vertex; all of them 
are available from lattice simulations, 
with the exception of a particular partial derivative,
denoted by $\w(r^2)$, related to the ghost-gluon kernel that appears in the STI~\cite{Aguilar:2020yni,Aguilar:2020uqw}.

The importance of the calculation put forth above becomes particularly transparent   
when an additional theoretical ingredient is taken into account.
Specifically, as will become clear in the main body of the article,
$\Cfat(r^2)$ {\it coincides} exactly with the
wave function amplitude of the massless bound state poles associated with the
three-gluon vertex~\cite{Aguilar:2011xe,Ibanez:2012zk,Aguilar:2017dco,Binosi:2017rwj}. 
Thus, the form of $\Cfat(r^2)$ is determined from an entirely different procedure,
namely as the solution of the BSE mentioned earlier. This solution, in turn,
serves as a benchmark of our analysis, in  the sense that  
signals emerging from the WI treatment are expected to
be qualitatively compatible with the $\Cfat(r^2)$ obtained from the BSE~\cite{Aguilar:2011xe,Ibanez:2012zk,Aguilar:2017dco,Binosi:2017rwj}.

Our numerical analysis reveals that the $\Cfat(r^2)$ constructed by putting together all the  
ingredients of the WI deviates markedly from zero, showing an impressive resemblance to the
results obtained from the corresponding BSE. 
On the average, the signal obtained is $3.1\sigma$
away from the null hypothesis value, $\Cfat(r^2)=0$, which corresponds to the absence of the
Schwinger mechanism. Moreover, for momenta $r \geq 2$~GeV the deviation of the signal
from $\Cfat(r^2)=0$ exceeds the $5\sigma$, owing to a characteristic peak 
of $\Cfat(r^2)$ in the vicinity of $2$~GeV, and to the fact that the 
error bars assigned to the lattice points get reduced as one moves away from the
deep infrared region.

Let us finally mention that the principal uncertainty associated with the WI 
determination originates from the computation of the function $\w(r^2)$, which is not available from
lattice simulations, and has been approximated by a truncated version of the SDE of the ghost-gluon kernel.
As was explained in~\cite{Aguilar:2020uqw}, the simulation of this
function on the lattice is theoretically conceivable, but practically rather cumbersome.

The article is organized as follows. 
In Sec.~\ref{sec:wi} we explain in an Abelian context 
how the presence of longitudinally coupled massless poles
modifies the form of the WI satisfied by the pole-free part of a vertex. 
In Sec.~\ref{sec:wi3g} we derive the corresponding WI
for the pole-free part of the three-gluon vertex, introducing the displacement function  $\Cfat(r^2)$. 
Then, in Sec.~\ref{sec:dislat} we express $\Cfat(r^2)$ in terms of the 
three-gluon form factor, the gluon propagator and its derivative, the ghost dressing function, 
and the function $\w(r^2)$. Next, in Sec.~\ref{sec:bse} we present the BSE determination of $\Cfat(r^2)$.
In Sec.~\ref{sec:numan} we use lattice inputs for the components 
of the WI in order to determine the form of $\Cfat(r^2)$, and compare it to the corresponding
result obtained from the BSE.
 Finally, in Sec.~\ref{sec:conc} we present our discussion and conclusions. 
 In addition, certain topics have been relegated to three Appendices:
 Appendix~\ref{sec:nbse} contains technical details of the BSE treatment,  
in Appendix~\ref{sec:cw} we discuss the SDE-based determination of $\w(r^2)$,
while in Appendix~\ref{sec:inp} we collect the fits employed in our numerical analysis.
  

\section{\label{sec:wi} Ward identities in the presence of massless poles}

In this section we focus on the modifications induced to the form of the WIs 
when the vertices involved contain longitudinally coupled massless poles, which is one of the
trademarks of the Schwinger mechanism at the level of the vertices. 

In general, the derivation of the WI from the corresponding Takahashi identity,
or, in general, from a given STI, involves a Taylor expansion around the contracting momentum~\cite{Aguilar:2016vin}.
In the case of a function of a single variable, $f((p+q)^2)$, such as a propagator, 
the Taylor expansion proceeds through the elementary formula ($q \to 0$)
\bea
f((p+q)^2) &=& f(p^2) + q^\alpha \, \frac{\partial f(p^2)}{\partial p^\alpha}   + {\cal O}(q^2)
\nonumber\\
&=& f(p^2) + 2 (q\cdot p) \, \frac{\partial f(p^2)}{\partial p^2}  + {\cal O}(q^2)\,.
\label{Taylprop}
\eea
For a function $f(q,r,p)$, with $q+r+p=0$, such as a three-particle vertex or kernel, 
the Taylor expansion  around $q=0$ (and $p=-r$) gives 
\bea 
f(q,r,p) = f(0,r,-r) + q^\alpha \left[\frac{\partial f(q,r,p)}{\partial q^\alpha}\right]_{q=0} + {\cal O}(q^2)\,.
\label{Taylf}
\eea
Note that if the $f(q,r,p) = -f(q,p,r)$, as happens in the case of the
term associated with the massless pole in the $q$ channel (see below), then $f(0,r,-r) = 0$.

In order to fix the ideas, we employ a vertex with reduced tensorial structure, 
which obeys an Abelian STI. In particular, we consider one of the typical vertices of the
background field method (BFM)~\cite{DeWitt:1967ub,Honerkamp:1972fd,tHooft:1971qjg,Kallosh:1974yh,Kluberg-Stern:1974nmx,Abbott:1980hw, Shore:1981mj,Abbott:1983zw}, 
namely the $B(q){\bar c}(r) c(p)$ vertex, where $B$ denotes the background gluon\footnote{Within the BFM, the gauge field $A^{a}_\alpha$ is decomposed as
  $A^{a}_\alpha = B^{a}_\alpha + Q^{a}_\alpha$, where $B^{a}_\alpha$ is the background field 
and $Q^{a}_\alpha$ is the quantum (fluctuating) field.} and  ${\bar c}$ ($c$) the anti-ghost (ghost)
fields. Due to the residual invariance of the action under background gauge transformations, 
this vertex satisfies an Abelian STI that relates it to the inverse ghost propagator. 
Specifically, suppressing the gauge coupling $g$ and the color factor $f^{abc}$, and 
denoting the remainder of the vertex by $\widetilde\Gamma_\alpha(q,r,p)$, we have~\cite{Aguilar:2006gr,Binosi:2009qm}
\be 
q^\alpha \widetilde\Gamma_\alpha(q,r,p) = {D}^{-1}(p^2) - {D}^{-1}(r^2)\,,
\label{STI1}
\ee
where the ghost propagator is given by \mbox{ $D^{ab}(q^2)= i \delta^{ab}D(q^2)$}.
Note that, at tree level, $\widetilde{\g}_{\!0}^\alpha(q,r,p)=(r-p)^\alpha$.

At this point we will assume that the Schwinger mechanism is inactive, such that the
form factors comprising $\widetilde\g_\alpha(q,r,p)$
do not contain poles. In that case, 
one may carry out the Taylor expansion of both sides of \1eq{STI1}
according to \2eqs{Taylprop}{Taylf}. Specifically, the left hand-side (l.h.s) of 
\1eq{STI1} yields 
\be
[{\rm l.h.s}] = q^\alpha \widetilde\Gamma_\alpha(0,r,-r) \, + {\cal O}(q^2)\,, 
\label{lhs}
\ee
while the right hand-side (r.h.s) is simply given by 
\be
[{\rm r.h.s}] = q^\alpha \frac{\partial {D}^{-1}(r^2)}{\partial r^\alpha} \, + {\cal O}(q^2) \,.
\label{rhs}
\ee
Then, equating the coefficients of the terms linear in $q^\alpha$ on both sides, one obtains the simple, QED-like relation~\cite{Aguilar:2016vin}
\be
\widetilde\Gamma_\alpha(0,r,-r) = \, \frac{\partial {D}^{-1}(r^2)}{\partial r^\alpha}\,.
\label{WInopole}
\ee
Finally, at the level of the single form factor comprising $\widetilde\Gamma_{\alpha}(0,r,-r)$, namely 
\be
\widetilde\Gamma_\alpha(0,r,-r) = \widetilde{\cal A}(r^2)  r_\alpha \,,
\label{Gff}
\ee
we obtain directly from \1eq{WInopole}~\cite{Aguilar:2016vin}
\be
\widetilde{\cal A}(r^2) = 2 \frac{\partial {D}^{-1}(r^2)}{\partial r^2} \,.
\label{AwD}
\ee

\begin{figure}[t]
\includegraphics[width=0.85\textwidth]{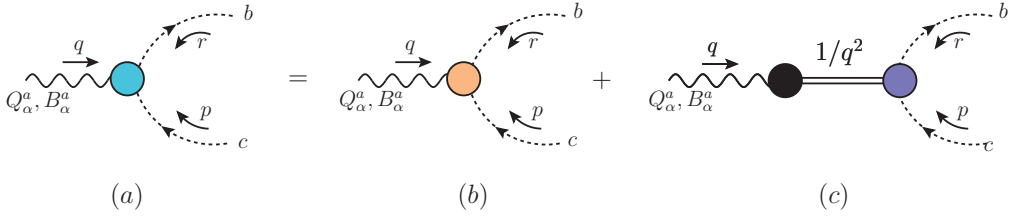} \\
\caption{ ($a$): Compact diagrammatic representation of the vertices $\fatgt_{\alpha}(q,r,p)$ ($B$ gluon)
  and $\fatg_{\alpha}(q,r,p)$ ($Q$ gluon); ($b$) the pole-free parts,
  $\widetilde\Gamma_{\alpha}(q,r,p)$ and $\Gamma_{\alpha}(q,r,p)$; ($c$) the pole parts,
  ${\widetilde V}_{\alpha}(q,r,p)$ and ${V}_{\alpha}(q,r,p)$. The black circle
  denotes the ``transition amplitude'', describing the mixing of the gluon with a massless excitation; 
its detailed diagrammatic content may be found in~\cite{Ibanez:2012zk}.} 
\label{fig:split_ggvpole}
\end{figure}

Let us now turn the Schwinger mechanism on, and denote the resulting full vertex by $\widetilde\fatg_\alpha(q,r,p)$. 
The vertex $\widetilde\fatg_\alpha(q,r,p)$,  diagrammatically represented in Fig.~\ref{fig:split_ggvpole},  is comprised by two distinct pieces, 
\be
\fatgt_\alpha(q,r,p) = \widetilde\Gamma_{\alpha}(q,r,p) + {\widetilde V}_{\alpha}(q,r,p),
\label{ghsm}
\ee
where $\widetilde\Gamma_{\alpha}(q,r,p)$ contains all pole-free contributions,
while the pole term ${\widetilde V}_{\alpha}(q,r,p)$ 
has the general form~\cite{Jackiw:1973tr,Cornwall:1973ts,Eichten:1974et,Poggio:1974qs,Smit:1974je}, 
\be
{\widetilde V}_{\alpha}(q,r,p)  = \frac{q_\alpha}{q^2}{\widetilde C}(q,r,p)\,.
\label{Vgh}
\ee
We emphasize that
the pole-free terms $\widetilde\Gamma_{\alpha}(q,r,p)$ 
are  {\it different} when the Schwinger mechanism is turned on or off. In particular,    
the infrared finiteness of the gluon propagator affects the behavior of all other Green's function,
due to its nontrivial interconnection with them imposed by the corresponding coupled SDEs. A typical
qualitative example of the type of modifications that the emergence of a gluonic mass scale
induces to one-loop contributions is the
conversion of ``unprotected'' logarithms into ``protected'' ones, according to
\mbox{$\ln({q^2}/{\mu^2})  \to \ln[(q^2+m^2)/{\mu^2}]$}~\cite{Aguilar:2013vaa,Athenodorou:2016oyh}. 

Evidently, combining \2eqs{ghsm}{Vgh} we get
\be
q^\alpha\fatgt_\alpha(q,r,p) = q^\alpha \widetilde\Gamma_{\alpha}(q,r,p) + \widetilde{C}(q,r,p)\,;
\label{contr}
\ee
thus, the contraction by $q^\alpha$ cancels the massless pole in $q^2$. 

We next assume that the Becchi-Rouet-Stora-Tyutin (BRST) symmetry~\cite{Becchi:1975nq,Tyutin:1975qk} of the theory 
remains intact as the Schwinger mechanism becomes operational.
In particular, the STIs satisfied by the elementary vertices are assumed to retain their standard form, but now   
being realized through the nontrivial participation of the massless pole terms~\mbox{\cite{Eichten:1974et,Poggio:1974qs,Smit:1974je,Cornwall:1981zr,Papavassiliou:1989zd,Aguilar:2008xm,Binosi:2012sj,Aguilar:2016vin}}. For a variety of perspectives  
related to the BRST symmetry in the presence of a mass gap, see,
\eg~\cite{Alkofer:2000wg,Fischer:2008uz,Alkofer:2011pe,Capri:2015ixa}, and references therein.

Accordingly, 
the full $\fatgt_\alpha(q,r,p)$ satisfies, as before, precisely \1eq{STI1}, namely
\be 
q^\alpha \fatgt_\alpha(q,r,p) = {D}^{-1}(p^2) - {D}^{-1}(r^2)\,,
\label{STI1sm}
\ee
where  $D(q^2)$ is the ghost propagator in the presence of the
Schwinger mechanism. For the same reasons described above for the case of $\widetilde\Gamma_{\alpha}(q,r,p)$, 
$D(q^2)$ also differs from the corresponding quantity
when the Schwinger mechanism is not operational.  

Then,  using \1eq{contr}, we obtain for the pole-free part 
\be
q^\alpha\widetilde\Gamma_{\alpha}(q,r,p) = \left[ D^{-1}(p^2) - D^{-1}(r^2)\right] \, - \widetilde{C}(q,r,p) \,.
\label{STI2}
\ee
The WI obeyed by $\widetilde\Gamma_{\alpha}(q,r,p)$ may be derived again by means 
of a Taylor expansion, since, after the contraction by $q^\alpha$, 
all terms appearing in the STI of \1eq{STI2} contain no poles, as $q\to 0$.
In particular, 
\be
q^\alpha \widetilde\Gamma_{\alpha}(0,r,-r) = \widetilde{C}(0,r,-r) + q^\alpha \left\{ \frac{\partial {D}^{-1}(r^2)}{\partial r^\alpha}
- \left[\frac{\partial \widetilde{C}(q,r,p)}{\partial q^\alpha}\right]_{q=0}\right\}  \,+ {\cal O}(q^2)\,.
\label{lhssm}
\ee
The comparison between \2eqs{lhssm}{rhs} reveals that the only zeroth-order contribution,
namely $\widetilde{C}(0,r,-r)$, must vanish,
\be
\widetilde{C}(0,r,-r) = 0 \,. 
\label{Cant}
\ee
Note that the result of \1eq{Cant} may be independently obtained from the property
$\widetilde{C}(q,r,p) = - \widetilde{C}(q,p,r)$, which follows directly from the general ghost-antighost symmetry of the
$B(q){\bar c}(r) c(p)$ vertex.

Then, the matching of the terms linear in $q$ yields the WI 
\be
\widetilde\Gamma_{\alpha}(0,r,-r)  = \frac{\partial {D}^{-1}(r^2)}{\partial r^\alpha} -
\underbrace{\left[\frac{\partial \widetilde{C}(q,r,p)}{\partial q^\alpha}\right]_{q=0}}_{\rm WI\,\, displacement} \,,
\label{WIdis}
\ee
which, when compared to that of \1eq{WInopole}, is ``displaced'' by 
the partial derivative of the form factor associated with the pole term. 

In order to determine the displaced analogue of \1eq{AwD}, we set
\be 
\left[\frac{\partial \widetilde{C}(q,r,p)}{\partial q^\alpha}\right]_{q = 0} \!\!\!\!\!= 
2 r_\alpha\,\Ctilde(r^2)\,, \qquad \Ctilde(r^2) := 
\left[ \frac{\partial \widetilde{C}(q,r,p)}{\partial p^2} \right]_{q = 0} \,,
\label{polegh}
\ee
and obtain immediately from \2eqs{Gff}{WIdis}
\be
\widetilde{\cal A}(r^2) = 2\left[ \frac{\partial {D}^{-1}(r^2)}{\partial r^2} -  \,\Ctilde(r^2)\right]\,.
\label{GDC}
\ee

Note that the displacement of the WI exemplified above becomes especially relevant 
within the framework that combines the pinch-technique (PT)~\cite{Cornwall:1981zr,Cornwall:1989gv,Pilaftsis:1996fh,Binosi:2009qm} with the BFM,
known as ``PT-BFM scheme''~\cite{Aguilar:2006gr,Binosi:2007pi}.
In particular, the action of terms such as $\Ctilde(r^2)$ is instrumental for
the evasion of a powerful nonperturbative cancellation that operates at the level of
the gluon SDE~\cite{Aguilar:2009ke},
which would otherwise enforce the result $\Delta^{-1}(0) = 0$.
In fact, the contribution of the ghost loop to the nonvanishing $\Delta^{-1}(0)$,
to be denoted by $\Delta_{\rm gh}^{-1}(0)$, is given by~\cite{Aguilar:2016vin}
\be
\Delta_{\rm gh}^{-1}(0) \sim \int\!\! d^4 k \, k^2 D^2(k^2) \Ctilde(k^2)\,.
\label{Dgh}
\ee
Let us finally point out that the displacement associated with the conventional ghost-gluon vertex $\fatg_\alpha(q,r,p)$
[see Fig.~\ref{fig:split_ggvpole}],
to be denoted by $\C(r^2)$ [see \1eq{polegh_conv}], is related to $\Ctilde(r^2)$ by the simple relation 
\be
\C(r^2) = F(0)\,\Ctilde(r^2) \,.
\label{cvsc}
\ee
The demonstration of \1eq{cvsc} relies on the ``background-quantum identity'' that relates $\widetilde\fatg_{\alpha}(q,r,p)$
and  $\fatg_\alpha(q,r,p)$~\cite{Binosi:2002ez,Binosi:2009qm}; details will be presented elsewhere.

\section{\label{sec:wi3g}Three-gluon vertex and its Ward identity displacement}

In this section we consider the case of the three-gluon vertex
in the conventional Landau gauge. If this vertex 
develops longitudinally coupled massless poles, its
pole-free part satisfies a displaced WI, whose derivation is the focal point of this section.

\begin{figure}[h]
\includegraphics[width=0.85\textwidth]{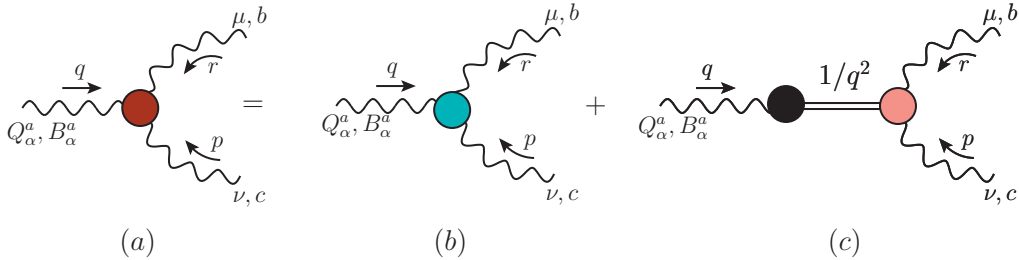} \\
\caption{($a$): Compact diagrammatic representation of the vertices $\fatgt_{\alpha\mu\nu}(q,r,p)$ ($B$ gluon)
  and $\fatg_{\alpha\mu\nu}(q,r,p)$ ($Q$ gluon); ($b$) the pole-free parts,
  ${\widetilde\g}_{\alpha\mu\nu}(q,r,p)$ and $\g_{\alpha\mu\nu}(q,r,p)$; ($c$) the pole parts,
  ${\widetilde V}_{\alpha\mu\nu}(q,r,p)$ and ${V}_{\alpha\mu\nu}(q,r,p)$.} 
\label{fig:split_3gpole}
\end{figure}

Before commencing, we introduce 
the gluon propagator, \mbox{$\Delta^{ab}_{\mu\nu}(q)=-i\delta^{ab}\Delta_{\mu\nu}(q)$}; in the {\it Landau gauge} that we employ in this work, it is given by the completely transverse form
\begin{align}
\Delta_{\mu\nu}(q) = \Delta(q^2) {P}_{\mu\nu}(q)\,, \qquad {P}_{\mu\nu}(q) := g_{\mu\nu} - q_\mu q_\nu/{q^2}\,.
\end{align}

In addition, we will use the ghost dressing function, $F(q^2)$, related to the ghost propagator by $F(q^2)= q^2 D(q^2)$. 
Furthermore, we define the two tensorial structures
\be
{P}_{\mu'}^{\mu}(r){P}_{\nu'}^{\nu}(-r) :={\cal T}_{\mu'\nu'}^{\mu\nu}(r)\,, \qquad \lambda_{\mu\nu\alpha}(r) := 2r_\alpha P_{\mu\nu}(r)\,,  
\label{defproj}
\ee
and the tree-level three-gluon vertex, \mbox{$\gz^{\alpha\mu\nu}(q,r,p)$}, as
\be 
\gz^{\alpha\mu\nu}(q,r,p) = ( q - r )^\nu g^{\alpha\mu} + ( r - p )^\alpha g^{\mu\nu} + ( p - q )^\mu g^{\nu\alpha} \,, 
\label{bare3g}
\ee
where the  gauge coupling $g$ and the color factor $f^{abc}$ were suppressed. 

In order for the Schwinger mechanism to be activated, 
the full three-gluon vertex, to be denoted by \mbox{$\fatg^{abc}_{\alpha\mu\nu}(q,r,p) = gf^{abc} \fatg_{\alpha\mu\nu}(q,r,p)$}, 
is written as (see Fig.~\ref{fig:split_3gpole})
\be 
\fatg_{\alpha\mu\nu}(q,r,p) = \g_{\alpha\mu\nu}(q,r,p) + V_{\alpha\mu\nu}(q,r,p) \,,
\label{fatG}
\ee
where $\g_{\alpha\mu\nu}(q,r,p)$ is the pole-free component, while 
$V_{\alpha\mu\nu}(q,r,p)$ contains {\it longitudinally coupled} poles, \ie\, it assumes the
general form 
\be 
V_{\alpha\mu\nu}(q,r,p) = \left( \frac{q_\alpha}{q^2} \right) C_{\mu\nu}(q,r,p) + \left( \frac{r_\mu}{r^2} \right) A_{\alpha\nu}(q,r,p) + \left( \frac{p_\nu}{p^2} \right) B_{\alpha\mu}(q,r,p) \,.
\label{eq:Vgen}
\ee

For the particular kinematic limit that we will eventually consider in the present work \mbox{($q\to 0$)},
we only require the tensorial decomposition
of the term  $C_{\mu\nu}(q,r,p)$ in \1eq{eq:Vgen}, given by 
\be 
C_{\mu\nu}(q,r,p) = C_1 \, g_{\mu\nu}  + C_2\, r_\mu r_\nu  + C_3 \, p_\mu p_\nu  +  C_4 \, r_\mu p_\nu  + C_5 \,  p_\mu r_\nu  \,,
\label{eq:Cdec}
\ee
where \mbox{$C_i := C_i(q,r,p)$}.

Due to its special form given by \1eq{eq:Vgen}, \mbox{$V_{\alpha\mu\nu}(q,r,p)$} satisfies the crucial condition  
\be
\label{eq:transvp}
{P}_{\alpha'}^{\alpha}(q){P}_{\mu'}^{\mu}(r){P}_{\nu'}^{\nu}(p) V_{\alpha\mu\nu}(q,r,p) = 0 \,,
\ee
and, consequently, it drops out from the typical lattice observables
involving the {\it transversely projected} three-gluon vertex [see \2eqs{asymlat}{gbar_def}].

The full vertex $\fatg_{\alpha\mu\nu}(q,r,p)$ satisfies the STI
\be
q^\alpha \fatg_{\alpha \mu \nu}(q,r,p) = F(q^2)
\left[\Delta^{-1}(p^2) P_\nu^\sigma(p) H_{\sigma\mu}(p,q,r) - \Delta^{-1}(r^2) P_\mu^\sigma(r) H_{\sigma\nu}(r,q,p)\right]\,; 
\label{stip}
\ee
analogous expressions are obtained when contracting by $r^\mu$ or $p^\nu$. 
Note that the ghost-gluon kernel, $H^{abc}_{\nu\mu}(q,p,r) = -gf^{abc}H_{\nu\mu}(q,p,r)$, defined in \fig{fig:HSDE}, enters in the STI nontrivially;
for the nonperturbative structure of its relevant form factors, see \cite{Aguilar:2018csq}.
In addition, we point out that the $H_{\sigma\mu}(p,q,r)$ and $H_{\sigma\nu}(r,q,p)$ also contain massless poles
in the $r_\mu$ and $p_\nu$ channels, respectively, which are completely eliminated by the transverse projections in \1eq{eq:stidef}.

It is clear from \2eqs{fatG}{eq:Cdec} that 
\be
{P}_{\mu'}^{\mu}(r){P}_{\nu'}^{\nu}(p) \left[q^\alpha \fatg_{\alpha \mu \nu}(q,r,p)\right] =  {P}_{\mu'}^{\mu}(r){P}_{\nu'}^{\nu}(p) 
[q^\alpha \Gamma_{\alpha \mu \nu}(q,r,p) +  C_{\mu\nu}(q,r,p)]\,,
\label{eq:stilhs} 
\ee
while, from the STI of \1eq{stip}
\be
{P}_{\mu'}^{\mu}(r){P}_{\nu'}^{\nu}(p) \left[q^\alpha \fatg_{\alpha \mu \nu}(q,r,p)\right]
= {P}_{\mu'}^{\mu}(r){P}_{\nu'}^{\nu}(p)\,F(q^2)\, {R}_{\nu\mu}(p,q,r)\,,
\label{eq:stirhs} 
\ee
where 
\be
{R}_{\nu\mu}(p,q,r) := \Delta^{-1}(p^2) H_{\nu\mu}(p,q,r) - \Delta^{-1}(r^2) H_{\mu\nu}(r,q,p)\,.
\label{eq:defR}
\ee
Then, equating the right-hand sides of \2eqs{eq:stilhs}{eq:stirhs} we obtain
\be
q^\alpha \left[{P}_{\mu'}^{\mu}(r){P}_{\nu'}^{\nu}(p) \Gamma_{\alpha \mu \nu}(q,r,p)\right] =
{P}_{\mu'}^{\mu}(r){P}_{\nu'}^{\nu}(p)   \left[ \,F(q^2)\, {R}_{\nu\mu}(p,q,r) - C_{\mu\nu}(q,r,p) \right]\,.
\label{eq:stidef}
\ee

Due to the presence of the projectors \mbox{${P}_{\mu'}^{\mu}(r){P}_{\nu'}^{\nu}(p)$}, 
it is clear from \1eq{eq:Cdec} that only the terms
\mbox{$C_1g_{\mu\nu}$} and \mbox{$C_5\, p_\mu r_\nu$} contribute to \mbox{$C_{\mu\nu}(q,r,p)$}. Note, however, that
since
\mbox{${P}_{\mu'}^{\mu}(r){P}_{\nu'}^{\nu}(p)  C_5 \,p_\mu r_\nu ={P}_{\mu'}^{\mu}(r){P}_{\nu'}^{\nu}(p)  C_5\,q_\mu q_\nu$}, this 
term is subleading, \ie of order \mbox{${\cal O}(q^2)$}, when the limit \mbox{$q \to 0$} is taken.

We next proceed with the implementation of the limit \mbox{$q \to 0$}. In particular, 
as was done in the previous section, we carry out the Taylor expansion of
both sides of \1eq{eq:stidef} around $q=0$, and collect terms linear in $q$.

The computation of the l.h.s. of \1eq{eq:stidef} is immediate: using \1eq{defproj}, we have   
\be
[{\rm l.h.s}] = q^\alpha {\cal T}_{\mu'\nu'}^{\mu\nu}(r) \Gamma_{\alpha \mu \nu}(0,r,-r)  + {\cal O}(q^2)\,.
\label{lhs3g}
\ee

The computation of the r.h.s. of \1eq{eq:stidef} is considerably more complicated.
We start by noticing that, to lowest order in $q$, 
only the term $C_1(q,r,p) g_{\mu\nu}$ survives. In addition, since 
it is clear from \1eq{eq:defR} that ${R}_{\nu\mu}(-r,0,r) = 0$,
the vanishing of the zeroth order contribution imposes the condition 
\be
C_1(0,r,-r) = 0 \,, \label{C1_0}
\ee
in exact analogy to \1eq{Cant}. 

Thus, the r.h.s. of \1eq{eq:stidef} becomes
\be
[{\rm r.h.s}] = q^\alpha\, {\cal T}_{\mu'\nu'}^{\mu\nu}(r)
F(0) \left[ \frac{\partial R_{\nu\mu}(p,q,r)}{\partial q^\alpha} \right]_{\! q = 0} \!\!\!\!\!
-  q^\alpha\, {P}_{\mu'\nu'}(r)\left[\frac{\partial C_{1}(q,r,p)}{\partial q^\alpha} \right]_{\!q = 0}  \!\!\!\!\! + {\cal O}(q^2)\,.
\label{eq:parder}
\ee

In order to compute the first partial derivative in \1eq{eq:parder}, we exploit the fact that, in the Landau gauge,   
the ghost-gluon kernel may be cast in the form~\cite{Ibanez:2012zk,Aguilar:2020yni} 
\be 
H_{\nu\mu}(p,q,r) = {\widetilde Z}_1 g_{\nu\mu} + q^\rho K_{\nu\mu\rho}(p,q,r) \,, \qquad
H_{\mu\nu}(r,q,p) = {\widetilde Z}_1 g_{\mu\nu} + q^\rho K_{\mu\nu\rho}(r,q,p)\,, 
\label{HtoK}
\ee
where the kernels $K$ do not contain poles as \mbox{$q\to 0$}.  Moreover,   ${\widetilde Z}_1$
is the {\it finite} constant renormalizing $H_{\nu\mu}(p,q,r)$ 
in the ``asymmetric'' momentum subtraction (MOM) scheme, employed in the lattice simulation of~\cite{Athenodorou:2016oyh,Aguilar:2019uob};
its numerical value, estimated in~\cite{Aguilar:2021okw}, is ${\widetilde Z}_1\approx 0.95$.

Then, to lowest order in $q$,   
\be
\left[\frac{\partial H_{\nu\mu}(p,q,r)}{\partial q^\alpha } \right]_{q=0} \!\!\!\!\!\!=  K_{\nu\mu\alpha}(-r,0,r)\,, \qquad
\left[\frac{\partial H_{\mu\nu}(r,q,p)}{\partial q^\alpha } \right]_{q=0}  \!\!\!\!\!\!= K_{\mu\nu\alpha}(r,0,-r)\,.
\label{Kdef1}
\ee

Consider next the tensor decomposition of $K_{\mu\nu\alpha}(r,0,-r)$~\cite{Aguilar:2020yni}, 
\be 
K_{\mu\nu\alpha}(r,0,-r) = \kw(r^2) g_{\mu\nu}r_\alpha + \cdots \,, 
\label{HKtens}
\ee
where the ellipses denote terms proportional to $g_{\nu\alpha}r_\mu$,  $g_{\mu\alpha}r_\nu$, and $r_\alpha r_\mu r_\nu$,
which get annihilated by contraction with ${\cal T}_{\mu'\nu'}^{\mu\nu}(r)$.
 Clearly, ${\cal T}_{\mu'\nu'}^{\mu\nu}(r)K_{\nu\mu\alpha}(-r,0,r) = - {\cal T}_{\mu'\nu'}^{\mu\nu}(r)K_{\mu\nu\alpha}(r,0,-r)$.

Then, it is straightforward to demonstrate that
\be
{\cal T}_{\mu'\nu'}^{\mu\nu}(r)\left[ \frac{\partial {R}_{\nu\mu}(p,q,r)}{\partial q^\alpha} \right]_{\! q = 0} \!\!\!\!\!\!= \lambda_{\mu'\nu'\alpha}(r)
\left\{\widetilde{Z}_1 [ \Delta^{-1}(r^2) ]^\prime - \kw(r^2)\Delta^{-1}(r^2) \right\} \,,
\label{oneside}
\ee
where the ``prime'' denotes differentiation with respect to $r^2$.

As for the second partial derivative in \1eq{eq:parder}, applying the chain rule we have 
\be 
\left[\frac{\partial C_1(q,r,p)}{\partial q^\alpha}\right]_{q = 0} \!\!\!\!\!\!= 2 r_\alpha \Cfat(r^2)\,, \qquad \Cfat(r^2) := \left[ \frac{\partial C_1(q,r,p)}{\partial p^2} \right]_{q = 0} \,,
\label{poleres}
\ee
such that 
\be   
P_{\mu'\nu'}(r) \left[ \frac{\partial C_{1}(q,r,p)}{\partial q^\alpha} \right]_{q = 0}\!\!\!\!\!\! = \lambda_{\mu'\nu'\alpha}(r) \Cfat(r^2)\,,
\label{otherside}
\ee
and, therefore, \1eq{eq:parder} becomes 
\be
   [{\rm r.h.s}] = q^\alpha \lambda_{\mu'\nu'\alpha}(r)\left[
F(0)\left\{\widetilde{Z}_1 [\Delta^{-1}(r^2)]^\prime - \kw(r^2) \Delta^{-1}(r^2)\right\}- \Cfat(r^2)\right]+ {\cal O}(q^2) \,.
\label{eq:PPGamma} 
\ee

The final step is to equate the terms linear in $q$ that appear in \2eqs{lhs3g}{eq:PPGamma}, to obtain
the WI 
\be
{\cal T}_{\mu'\nu'}^{\mu\nu}(r) \Gamma_{\alpha \mu \nu}(0,r,-r) = 
\lambda_{\mu'\nu'\alpha}(r)\left[
F(0)\left\{\widetilde{Z}_1 [\Delta^{-1}(r^2)]^\prime - \kw(r^2) \Delta^{-1}(r^2)\right\}- \Cfat(r^2)\right]\,.
\label{WIdis3g} 
\ee

Thus, the inclusion of the term  $V_{\alpha\mu\nu}(q,r,p)$ in the vertex
of \1eq{fatG} leads ultimately to the displacement of the WI satisfied by
the pole-free part $\Gamma_{\alpha \mu \nu}(q,r,p)$, by an amount given by the 
special function $\Cfat(r^2)$. 
Evidently, if $\Cfat(r^2)=0$ one recovers the WI in the absence of the Schwinger mechanism. 

We end this section with some remarks related to the PT-BFM scheme.
Note that if the gauge field carrying the momentum $q$ is a background gluon instead of a quantum one (see Fig.~\ref{fig:split_3gpole}), 
then the corresponding three-gluon vertex, $\widetilde\Gamma_{\alpha \mu \nu}(q,r,p)$, satisfies 
a simplified version of the WI in \1eq{WIdis3g}, where $F\to 1$, $\widetilde{Z}_1 \to 1$, and $\kw(r^2)\to 0$, \ie 
\be
{\cal T}_{\mu'\nu'}^{\mu\nu}(r) \widetilde\Gamma_{\alpha \mu \nu}(0,r,-r) = 
\lambda_{\mu'\nu'\alpha}(r)\left[ [\Delta^{-1}(r^2)]^\prime - \widetilde\Cfat(r^2)\right]\,.
\label{WIdis3gbfm} 
\ee

As has been demonstrated in~\cite{Aguilar:2016vin},  
the contribution of the gluon loops to the nonvanishing $\Delta^{-1}(0)$,
to be denoted by $\Delta_{\rm gl}^{-1}(0)$, is controlled by $\widetilde\Cfat(r^2)$, 
\be
\Delta_{\rm gl}^{-1}(0) \sim \int\!\! d^4 k \, k^2 \Delta^2(k^2) \left[ 1 - 6\pi\alpha_s C_{\rm A}Y(k^2)\right] \Cfattilde(k^2)\,,
\label{Dgl}
\ee
where $\alpha_s := g^2/4\pi$,  $C_\mathrm{A}$ is the Casimir eigenvalue of the adjoint representation [$N$ for $SU(N)$], and $Y(k^2)$ represents
a particular one-loop correction (see, \eg Fig.~3 in ~\cite{Aguilar:2016vin}). Evidently, \1eq{Dgl} is the exact analogue of \1eq{Dgh}.
The total mass, identified with $\Delta^{-1}(0)$, is obtained by summing up \2eqs{Dgl}{Dgh}. 

Finally, the relation between $\widetilde\Cfat(r^2)$ and $\Cfat(r^2)$ is given by
\be
\Cfat(r^2) = F(0)\,\Cfattilde(r^2)  \,,
\label{gvsg}
\ee
in exact analogy to \1eq{cvsc}.

\section{\label{sec:dislat}Displacement function in terms of lattice quantities}

In this section we establish a crucial connection between the l.h.s. of \1eq{WIdis3g} and the
results of recent lattice simulations. This, in turn, will allow us to relate
the characteristic ingredient of the Schwinger mechanism, namely $\Cfat(r^2)$, to
quantities obtained directly from lattice QCD. The advantage
of such a connection is that the lattice is intrinsically ``blind'' to particular
field theoretic constructs (such as the Schwinger mechanism), furnishing results
obtained through the model-independent functional averaging over gauge-field configurations.

We start our analysis by considering 
the pole-free part $\Gamma_{\alpha\mu\nu}(q,r,p)$ of the three-gluon vertex, in the kinematic limit of interest, $q\to 0$.
Given that only a single momentum ($r$) is available, 
the general tensorial decomposition of $\Gamma_{\alpha\mu\nu}(0,r,-r)$ is given by  
\be
\Gamma_{\alpha\mu\nu}(0,r,-r) = 2 {\cal A}_1(r^2) \,r_\alpha g_{\mu\nu} + {\cal A}_2(r^2) (r_\mu g_{\nu\alpha} + r_\nu g_{\mu\alpha}) 
+ {\cal A}_3(r^2)\, r_\alpha r_\mu r_\nu \,,   
\label{Gtens}
\ee
where the form factors ${\cal A}_i(r^2)$ may diverge at most logarithmically as $r \to 0$, but do not contain
stronger singularities.  At tree level, we have that 
\be
\gz^{\alpha\mu\nu}(0,r,-r) = 2 \,r^\alpha g^{\mu\nu} - (r^\mu g^{\nu\alpha} + r^\nu g^{\mu\alpha}) \,, 
\label{G0tens}
\ee
corresponding to ${\cal A}_1^{(0)}(r^2)= 1$,  ${\cal A}_2^{(0)}(r^2) =-1$, and ${\cal A}_3^{(0)}(r^2)= 0$.

It is then elementary to derive from \1eq{Gtens} that
\be
{\cal T}_{\mu'\nu'}^{\mu\nu}(r) \Gamma_{\alpha\mu\nu}(0,r,-r) = {\cal A}_1(r^2) \lambda_{\mu'\nu'\alpha}(r)\,.
\label{TGamma}
\ee

We next establish a connection between the form factor ${\cal A}_1(r^2)$ and 
the projection of the three-gluon vertex studied in the lattice simulations of~\mbox{\cite{Parrinello:1994wd,Alles:1996ka,Parrinello:1997wm,Boucaud:1998bq,Cucchieri:2006tf,Cucchieri:2008qm,Duarte:2016ieu,Sternbeck:2017ntv,Vujinovic:2018nqc,Boucaud:2018xup,Aguilar:2019uob, Aguilar:2021lke}}.  Specifically, 
after appropriate amputation of the external legs, the lattice quantity $\Ls(r^2)$ is given by 
\bea
\Ls(r^2) &=&  \frac{\gz^{\alpha\mu \nu}(q,r,p)
P_{\alpha\alpha'}(q)P_{\mu\mu'}(r)P_{\nu\nu'}(p) \fatg^{\alpha'\mu'\nu'}(q,r,p)}
{\rule[0cm]{0cm}{0.45cm}\; {\gz^{\alpha\mu\nu}(q,r,p) P_{\alpha\alpha'}(q)P_{\mu\mu'}(r)P_{\nu\nu'}(p) \gz^{\alpha'\mu'\nu'}(q,r,p)}}
\rule[0cm]{0cm}{0.5cm} \Bigg|_{\substack{\!\!q\to 0 \\ p\to -r}} \,. 
\label{asymlat}
\eea
Now, by virtue of \1eq{eq:transvp}, it is clear that 
the term $V^{\alpha'\mu'\nu'}(q,r,p)$ associated with the poles drops out from \1eq{asymlat} in its entirety,
amounting effectively to the replacement $\fatg^{\alpha'\mu'\nu'}(q,r,p) \to \Gamma^{\alpha'\mu'\nu'}(q,r,p)$.

Then, the numerator, ${\cal N}$, and denominator, ${\cal D}$,
of the fraction on the r.h.s. of \1eq{asymlat}, after employing
\2eqs{Gtens}{G0tens}, become
\be
{\cal N} = 4 (d-1) [r^2 - (r\cdot q)^2/q^2] {\cal A}_1(r^2)\,, \qquad  {\cal D} = 4 (d-1) [r^2 - (r\cdot q)^2/q^2] \,.
\label{NandD}
\ee
Evidently, the path-dependent contribution contained in the square bracket drops out when forming the ratio ${\cal N}/{\cal D}$, and \1eq{asymlat} yields the important relation
\be
\Ls(r^2) =  {\cal A}_1(r^2) \,.
\label{LisB}
\ee

Combining \2eqs{TGamma}{LisB},  we get 
\be
{\cal T}_{\mu'\nu'}^{\mu\nu}(r)\Gamma_{\alpha\mu\nu}(0,r,-r) =  \Ls(r^2) \lambda_{\mu'\nu'\alpha}(r) \,.
\label{eq:PPG} 
\ee

At this point, after substitution of \1eq{eq:PPG} into \1eq{WIdis3g}, we arrive at 
\be
\Cfat(r^2) = F(0)\left\{\widetilde{Z}_1 [\Delta^{-1}(r^2)]^\prime - \kw(r^2) \Delta^{-1}(r^2)\right\}  - \Ls(r^2) \,. 
\label{centmink}
\ee

The final step consists in passing the result of \1eq{centmink} from Minkowski to Euclidean space, following the
standard conversion rules. Specifically, we set $r^2 = - r_{\s {\rm E}}^2$, with $r^2_{\s {\rm E}} > 0$ the positive square of an Euclidean four-vector, and use 
\bea
 \Delta_{\s {\rm E}}(r^2_{\s {\rm E}}) =& - \Delta(-r^2_{\s {\rm E}}) \,, \qquad F_{\s {\rm E}}(r^2_{\s {\rm E}}) =& F(-r^2_{\s {\rm E}}) \,,  \nonumber\\
 \Ls^{\s {\rm E}}(r^2_{\s {\rm E}}) =& \Ls(-r^2_{\s {\rm E}}) \,, \qquad \Cfat_{\s {\rm E}}(r^2_{\s {\rm E}}) =& - \Cfat(-r^2_{\s {\rm E}}) \,. 
\eea
In what follows we suppress the indices ``${\rm E}$'' to avoid notational clutter.

Then, \1eq{centmink} is converted to  
\be
\Cfat(r^2) = \Ls(r^2) + F(0)\left\{\kw(r^2)\Delta^{-1}(r^2) - \widetilde{Z}_1 [\Delta^{-1}(r^2)]^\prime \right\} \,,
\label{centeuck}
\ee
which is one of the central results of this article. 

Finally, it is convenient to introduce the dimensionless function $\w(r^2)$,
defined as 
\be
\kw(r^2) = - \frac{\w(r^2)}{r^2}\,,
\label{KW}
\ee
thus casting \1eq{centeuc} into the form
\be
\Cfat(r^2) = \Ls(r^2) - F(0)\left\{\frac{\w(r^2)}{r^2}\Delta^{-1}(r^2) + \widetilde{Z}_1 [\Delta^{-1}(r^2)]^\prime \right\} \,,
\label{centeuc}
\ee
which will be employed in the numerical evaluation given in Sec.~\ref{sec:numan}.

\section{\label{sec:bse} Dynamical determination of the displacement function}

In this section we elaborate on the determination of $\Cfat(r^2)$  from the BSEs that 
describe the dynamical formation of massless colored bound states;
the analysis is based on the derivations given in~\cite{Aguilar:2011xe,Aguilar:2017dco}, adapted to the present context.
Note that this procedure determines also the analogue of $\Ctilde(r^2)$, introduced in Sec.~\ref{sec:wi}, 
for the case of the conventional ghost-gluon vertex, to be denoted by $\C(r^2)$.

\begin{figure}[t]
\includegraphics[width=0.85\textwidth]{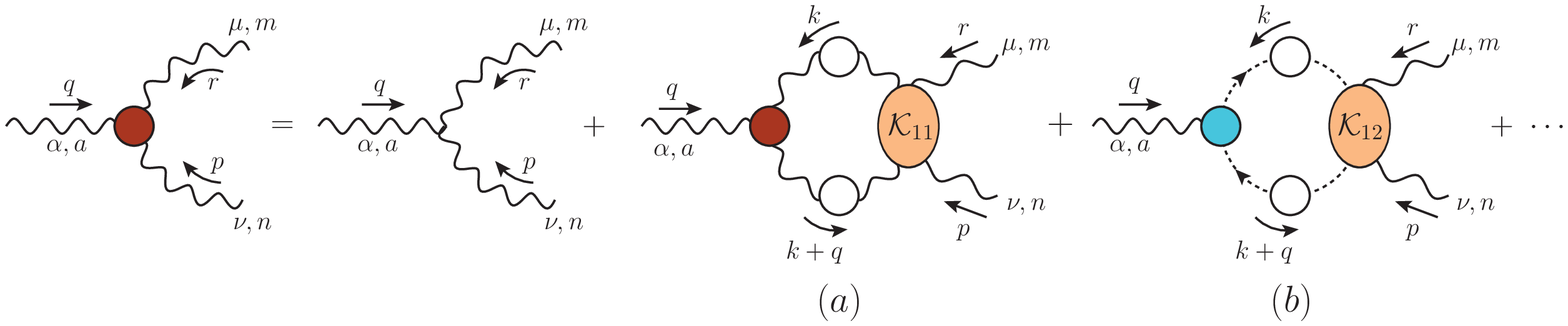} \\
\includegraphics[width=0.85\textwidth]{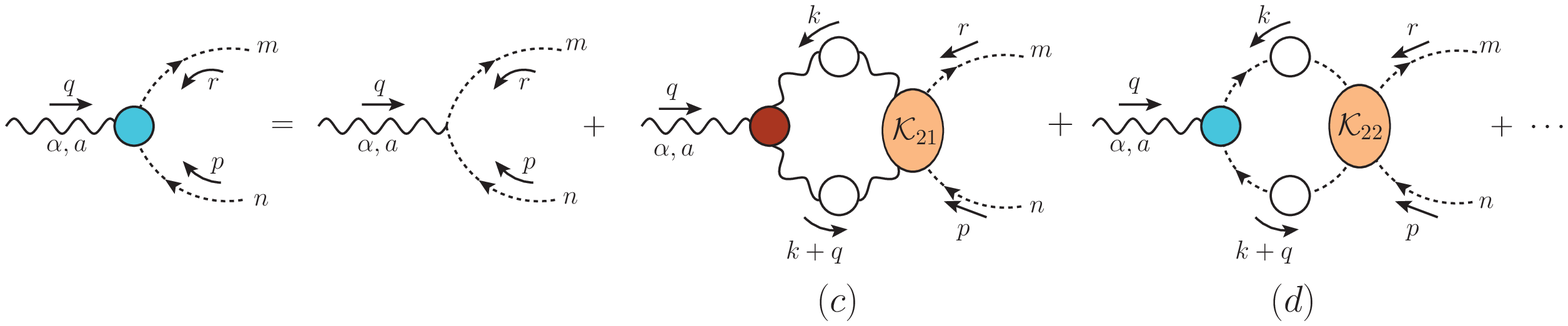}	
 \caption{ The coupled system of inhomogeneous BSEs for the three-gluon and ghost-gluon vertices, $\fatg^{\alpha\mu\nu}(q,r,p)$ and $\fatg^{\alpha}(q,r,p)$, respectively. Circles represent full propagators or vertices, while the orange ellipses correspond to four-point scattering kernels, ${\cal K}_{ij}$. The omitted  diagrams contain five-point scattering kernels,
associated with fully-dressed four-gluon vertices with incoming momentum $q$, which are neglected in our truncation scheme. }
\label{fig:BSEs_inhom}
\end{figure}

The starting point of this study is the BS version of the SDEs that govern the momentum evolution of the three-gluon vertex,
 $\fatg_{\alpha\mu\nu}(q,r,p)$, and of the conventional ghost-gluon vertex, $\fatg_\alpha(q,r,p)$, 
shown in \fig{fig:BSEs_inhom}. In particular, we replace (inside the loops) the
tree-level vertices (with incoming momentum $q$) by their fully-dressed counterparts, modifying
the corresponding multiparticle kernels ${\cal K}_{ij}$ accordingly, to avoid overcounting
(see \eg Fig.~7 of~\cite{Aguilar:2011xe}). 
The main advantage of this conversion is that various vertex renormalization constants, which otherwise
  would appear explicitly multiplying the corresponding diagrams, are naturally absorbed by the additional dressed vertices. 
Note that, in order to simplify the pertinent set of SDEs, we omit from our analysis the fully-dressed four-gluon vertices
 (with incoming momentum $q$), whose impact is expected to be subleading~\cite{Williams:2015cvx,Huber:2018ned}.

In what follows, we will introduce a longitudinally coupled massless pole also in 
  the ghost-gluon vertex $\fatg_\alpha(q,r,p)$,  casting it into a form analogous to \1eq{ghsm},  and diagrammatically
  represented in Fig.~\ref{fig:split_ggvpole}, where the incoming gluon is $Q^a_\alpha$ .  In particular, we set 
\be
\fatg_\alpha(q,r,p) = \g_{\alpha}(q,r,p) + V_{\alpha}(q,r,p),
\label{ghsm_conv}
\ee
where $\g_\alpha(r,p,q)$ denotes the pole-free component, while
\be 
V_\alpha(q,r,p) = \frac{q_\alpha}{q^2}\Cgh(q,r,p) \,,
\ee
describes the pole multiplied by the associated form factor.

Then, the BSEs of \fig{fig:BSEs_inhom} may be written schematically as
\bea
\fatg^{\alpha\mu\nu}(q,r,p) &=& \gz^{\alpha\mu\nu}(q,r,p) - \frac{i g^2 C_{\rm A}}{2} \int_k \fatg^{\alpha\beta\gamma}(q,k,-s)\Delta_{\beta\rho}(k)\Delta_{\gamma\sigma}(s){\cal K}_{11}^{\mu\nu\sigma\rho}(r,p,s,-k) \nonumber\\
&& + i g^2 C_{\rm A} \int_k \fatg^{\alpha}(q,k,-s)D(k^2)D(s^2){\cal K}_{12}^{\mu\nu}(r,p,s,-k) \,, \nonumber\\
\fatg^{\alpha}(q,r,p) &=& \gz^\alpha(q,r,p) - \frac{i g^2 C_{\rm A}}{2} \int_k \fatg^{\alpha\beta\gamma}(q,k,-s)\Delta_{\beta\rho}(k)\Delta_{\gamma\sigma}(s){\cal K}_{21}^{\sigma\rho}(r,p,s,-k) \nonumber\\
&& - \frac{i g^2 C_{\rm A}}{2} \int_k \fatg^{\alpha}(q,k,-s)D(k^2)D(s^2){\cal K}_{22}(r,p,s,-k) \,,
\label{BSE_inhom}
\eea
where $\gz^\alpha(q,r,p) = r^\alpha$ is the tree level ghost-gluon vertex, $s := k + q$, and we introduce the notation
\begin{equation}
\int_k := \frac{1}{(2\pi)^4} \int\!\! d^4 k \,, 
\label{intmeas}
\end{equation}
for the integral measure. In \1eq{intmeas}, the use of a symmetry-preserving regularization scheme is implicitly assumed.

  Next, we decompose the vertices in \1eq{BSE_inhom} according to \2eqs{fatG}{ghsm_conv}. Given that $\fatg_{\alpha\mu\nu}(q,r,p)$ contains poles in all channels, $q_\alpha$, $r_\mu$, and $p_\nu$, we isolate the pole in $q_\alpha$ by contracting
the first line of \1eq{BSE_inhom} with $P^{\mu\delta}(r)P^{\nu}_\delta(p)$. Then, using \2eqs{eq:Vgen}{eq:Cdec} we find that
\be 
P^{\mu\delta}(r)P^{\nu}_\delta(p)V_{\alpha\mu\nu}(q,r,p) = \frac{q_\alpha}{q^2}P^{\mu\delta}(r)P^{\nu}_\delta(p)\left[ g_{\mu\nu} C_1(q,r,p) + q_\mu q_\nu C_5(q,r,p) \right] \,, 
\label{PPV} 
\ee
so that the only pole terms on the l.h.s. of \1eq{BSE_inhom} are those containing $C_1$ and $C_5$. In addition, due to the transversality of the Landau gauge gluon propagator, \1eq{PPV} [with $q\to q$, $r\to k$, and $p\to -s$] can be used inside the integral of diagram ($a$);
again, only $C_1(q,k,-s)$ and $C_5(q,k,-s)$ survive. Exactly the same situation is reproduced inside diagram ($c$).

The following step is to multiply \1eq{BSE_inhom} by $q^2$ and expand around $q=0$.
In doing so, we recall \2eqs{C1_0}{poleres}, and the analogous relations for $\Cgh(q,r,p)$, namely\footnote{\1eq{Cant_conv} can be proved from an STI for the ghost-gluon vertex, in analogy to the proof leading to \1eq{Cant} using the Abelian STI of \1eq{contr}. The full derivation will be given elsewhere.}
\be 
\Cgh(0,r,-r) = 0 \,, 
\label{Cant_conv}
\ee
and
\be 
\left[\frac{\partial \Cgh(q,r,p)}{\partial q^\alpha}\right]_{q = 0} \!\!\!\!\!= 2 r_\alpha\,\C(r^2)\,, \qquad \C(r^2) := \left[ \frac{\partial \Cgh(q,r,p)}{\partial p^2} \right]_{q = 0} \!\!\!\!\!  \,.
\label{polegh_conv}
\ee
Then, as \mbox{$q \to 0$}, the term proportional to $C_5$ in \1eq{PPV}  is of higher order in $q$ and drops out.

\begin{figure}[t]
\includegraphics[width=0.75\textwidth]{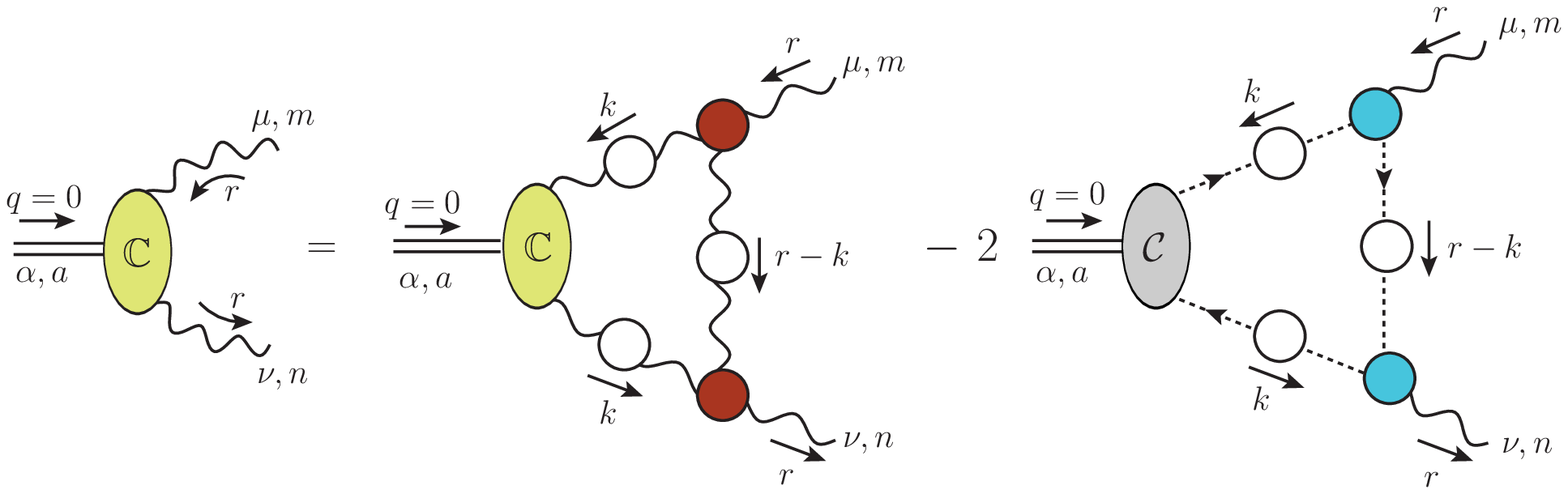} \\
\includegraphics[width=0.75\textwidth]{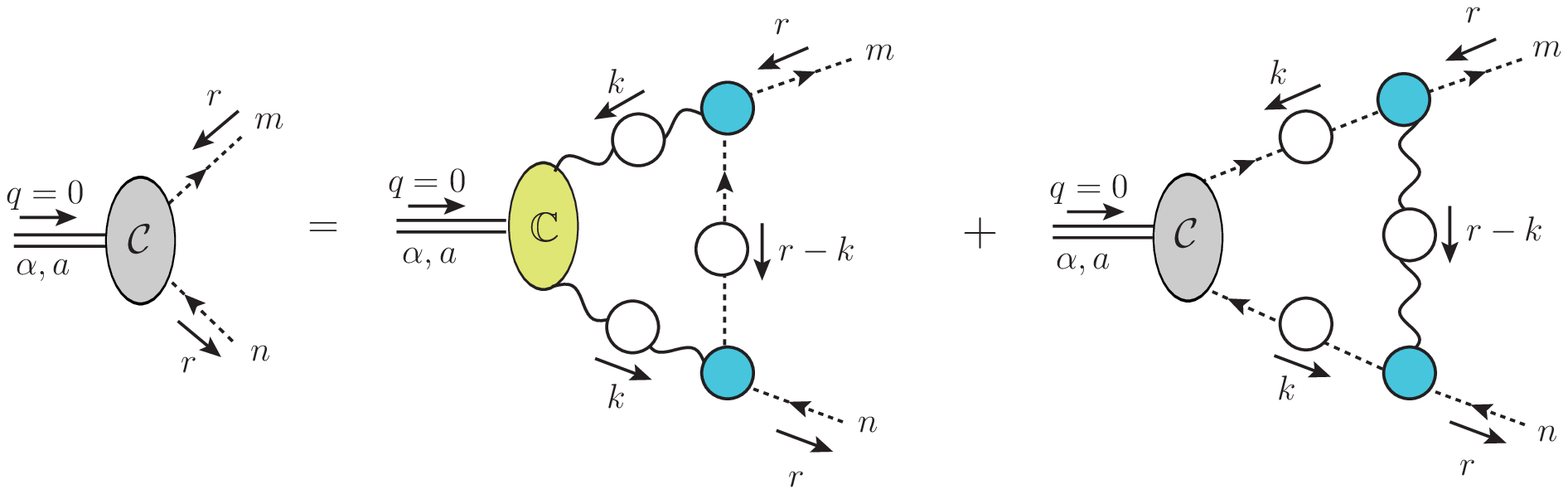}	
\caption{ The coupled system of BSEs governing the evolution of $\Cfat(r^2)$ and $\C(r^2)$. }
\label{fig:BSEs}
\end{figure}

Consequently, we obtain a set of homogeneous equations involving only $\Cfat(r^2)$ and $\C(r^2)$.
Specifically, we find  
\bea
3 (q\cdot r)\Cfat(r^2) &=& - \frac{i g^2 C_{\rm A}}{2}\int_k ( q\cdot k)\Cfat(k^2) \Delta^2(k^2)P_{\rho\sigma}(k)P_{\mu\nu}(r){\cal K}_{11}^{\mu\nu\sigma\rho}(r,-r,k,-k) \nonumber\\
&& + i g^2 C_{\rm A}\int_k (q\cdot k) \,\C(k^2) D^2(k^2)P_{\mu\nu}(r){\cal K}_{12}^{\mu\nu}(r,-r,k,-k) \,, \nonumber\\
(q\cdot r)\,\C(r^2) &=& - \frac{i g^2 C_{\rm A}}{2} \int_k (q\cdot k)\,\Cfat(k^2)\Delta^2(k^2)P_{\sigma\rho}(k){\cal K}_{21}^{\sigma\rho}(r,-r,k,-k) \nonumber\\
&& - \frac{i g^2 C_{\rm A}}{2} \int_k (q\cdot k) \,\C(k^2)D^2(k^2){\cal K}_{22}(r,-r,k,-k) \,, 
\label{BSE_hom}
\eea
where we have used $P^\mu_\mu(r)=3$. Finally, the remaining common factor of $q$ can be eliminated straightforwardly, by making use of 
the basic formula
\be
\int_k (q\cdot k) \,F(k,r) = \frac{(q\cdot r)}{r^2} \int_k (r\cdot k) \,F(k,r)\,.
\ee

Finally, we approximate
the four-point scattering kernels ${\cal K}_{ij}$ by their one-particle exchange diagrams (see \eg Figs.~4 and 5 of \cite{Aguilar:2017dco}), thus 
reducing the BSEs governing $\Cfat(r^2)$ and $\C(r^2)$ to the form shown in \fig{fig:BSEs}; the corresponding algebraic expressions
are given in \1eq{CsBSE_Euc}.

We observe that the system of integral equations reached in \1eq{BSE_hom} is the (approximate) 
BSE that governs the formation of massless colored bound states ($q^2=0$),
as announced\footnote{Note that the BSE derived as $q\to 0$ is identical to the one obtained as $q^2\to 0$;
however, the former derivation is operationally simpler.}.  
Thus, the function $\Cfat(r^2)$, connected with the displacement of the WI in \1eq{centeuc}, 
emerges{ \it naturally} as the wave function associated with the
pole formation of a colored two-gluon bound state.

We point out that, in addition to the lattice propagators given in Appendix~\ref{sec:inp}, the numerical evaluation of the BSEs 
requires information on various form factors of the pole-free vertices $\Gamma_{\alpha\mu\nu}(q,r,p)$ and  $\Gamma_\alpha(q,r,p)$;
for details, see Appendix~\ref{sec:nbse}.

\begin{figure}[t]
\includegraphics[width=0.475\textwidth]{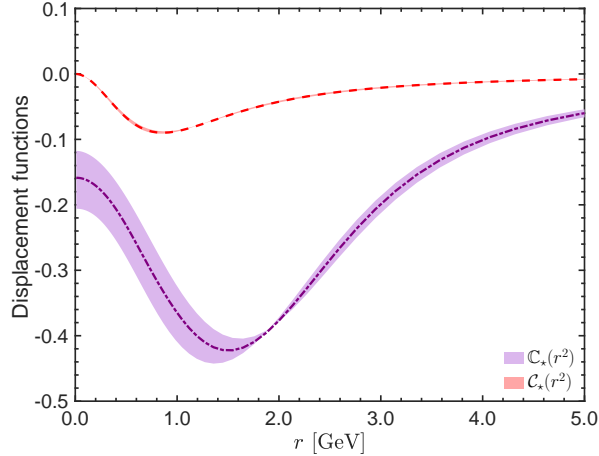}
\caption{ The solutions for $\CB(r^2)$ (purple dot-dashed) and $\Cc(r^2)$ (red dashed) obtained from the coupled BSE system of \1eq{CsBSE_Euc}.  
  The origin of the error bands is the propagation of the error associated with $\Ls(r^2)$,
used to calibrate the three-gluon vertex inputs.
 The band of $\Cc(r^2)$ turns out to be very thin and barely visible. }
\label{fig:C_gl_Cgh}
\end{figure}

We emphasize that, due to the homogeneity and linearity of \1eq{CsBSE_Euc}, the overall scale of the solution is undetermined:
the multiplication of a given solution  by an arbitrary real constant yields another solution.
For the purposes of the present work, this ambiguity was resolved by matching the
BSE prediction for $\Cfat(r^2)$ to the result obtained from the WI in the next section.
The solutions found for $\Cfat(r^2)$ and $\C(r^2)$ after the implementation of this scale-fixing procedure, 
denoted as $\CB(r^2)$ and $\Cc(r^2)$, respectively, are  
shown in \fig{fig:C_gl_Cgh}. Note that $\CB(r^2)$ is considerably larger in magnitude than $\Cc(r^2)$,
in agreement with the original study presented in~\cite{Aguilar:2017dco}.

\section{\label{sec:numan}  Displacement function from the Ward identity}

\begin{figure}[t]
\includegraphics[width=0.475\textwidth]{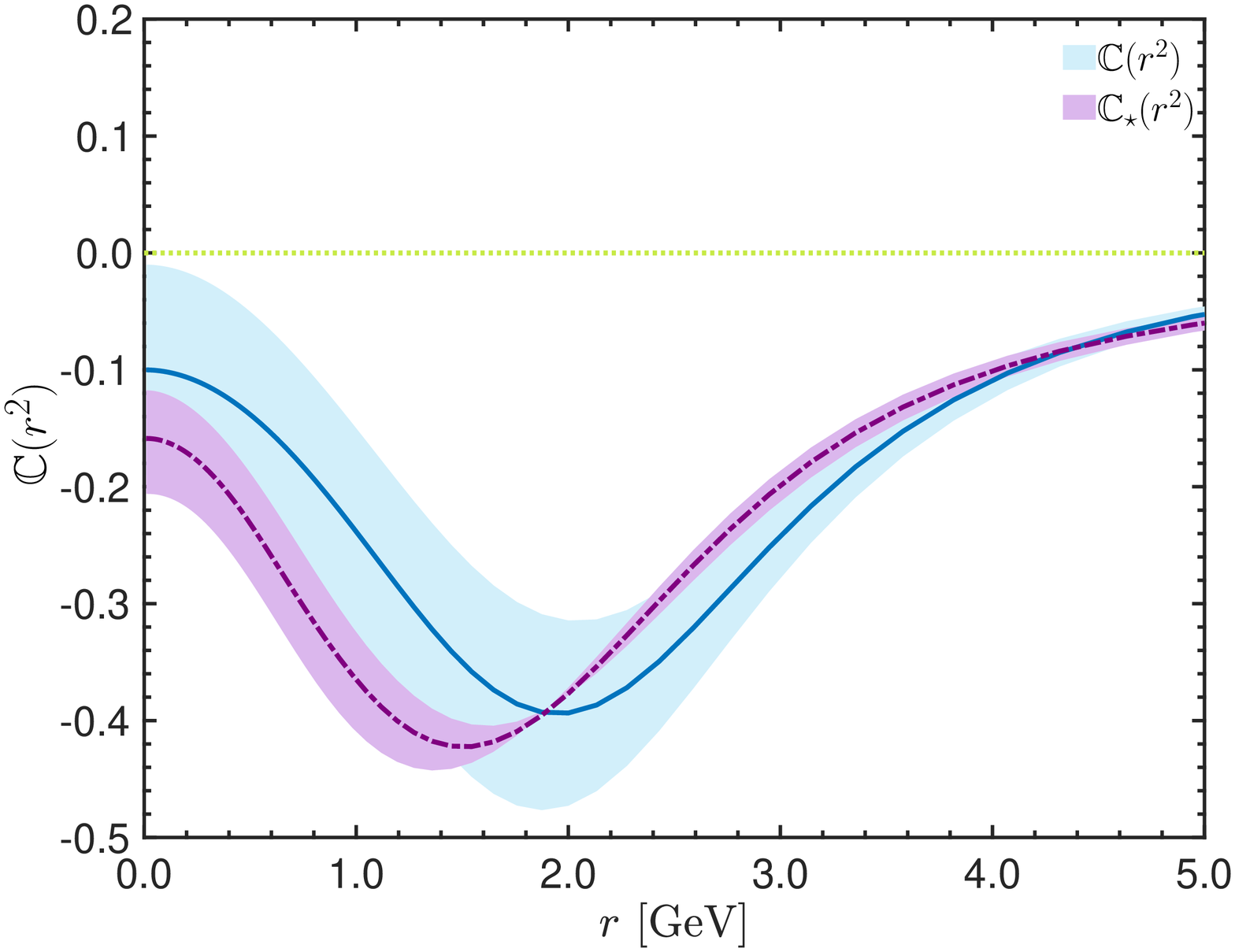} \hfill
\includegraphics[width=0.475\textwidth]{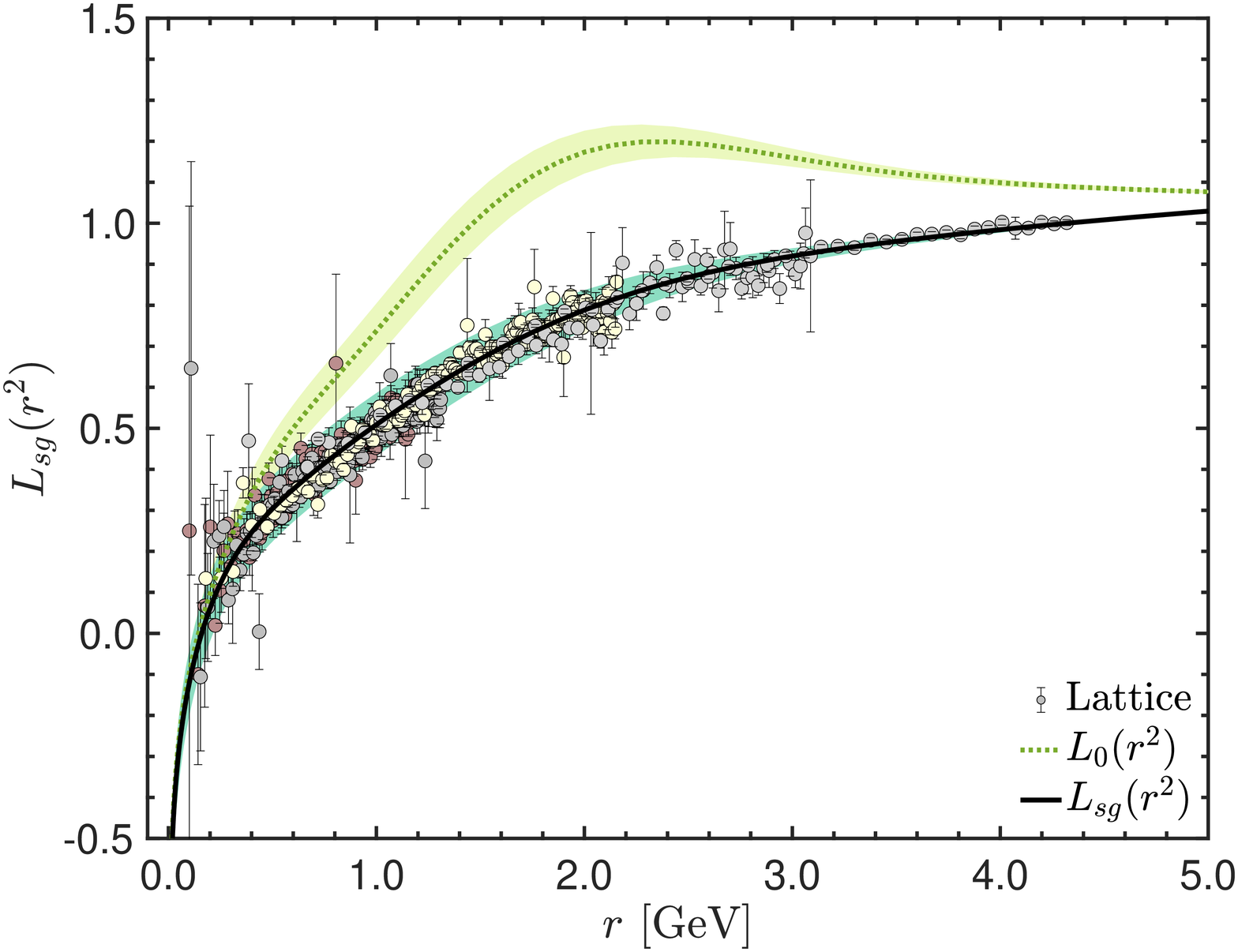}
\caption{ Left: $\Cfat(r^2)$ obtained from \1eq{centeuc} (blue continuous curve) compared to the BSE prediction, $\CB(r^2)$, based on \1eq{CsBSE_Euc} (purple dot-dashed). Right: Lattice data of \cite{Aguilar:2021lke} for $\Ls(r^2)$ (points), compared to the fit of \1eq{L_fit} (black continuous), and the null hypothesis prediction, $\Ln(r^2)$, of \1eq{L0} (green dotted).  The band around $\Ls(r^2)$ is delimited by the functions $\Ls^{\pm}(r^2)$ of \1eq{Lerr}.}
\label{fig:Cres}
\end{figure}

We next determine the signal for $\Cfat(r^2)$
that emerges from the corresponding WI, and discuss its statistical significance
with respect to the null hypothesis, namely the case where $\Cfat(r^2)$
would vanish identically. 

To that end, we substitute on the r.h.s. of \1eq{centeuc}
appropriate expressions for all quantities appearing there. 
In particular, we employ physically motivated fits to  
lattice results for the gluon propagator, the ghost dressing function, and $\Ls(r^2)$, 
given in Appendix~\ref{sec:inp}. Instead, the function $\w(r^2)$ 
is computed from its own SDE, as described in Appendix~\ref{sec:cw}; the resulting $\w(r^2)$ is shown as the blue solid line and error band in the right panel of \fig{fig:intgnd_Ws}.

The outcome of this operation is clearly nonvanishing: 
the resulting $\Cfat(r^2)$ is shown 
in the left panel of \fig{fig:Cres} as the blue continuous curve, which is distinctly 
separated from the null hypothesis case, indicated by the green dotted line.
The blue band surrounding the central result indicates 
the errors assigned to $\Cfat(r^2)$, through the propagation of the
corresponding errors associated with the ingredients entering on the r.h.s. of \1eq{centeuc}.

In the same figure we plot the $\CB(r^2)$ of \fig{fig:C_gl_Cgh}, in order to facilitate the direct comparison. 
We observe an excellent agreement in the overall shapes of $\CB(r^2)$ and $\Cfat(r^2)$. 
Their main difference is the position and depth of the minimum:
for $\Cfat(r^2)$ we have $r_{\s{ \rm min}} = 1.95 \substack{+ 0.07 \\ - 0.10}\,$ and 
\mbox{$\Cfat(r_{\s{ \rm min}}^2) = - 0.39 \pm 0.08$}, while for $\CB(r^2)$
we find $r_{\s{ \rm min}} = 1.50 \substack{+ 0.12 \\ - 0.13}\,$ and $\CB(r_{\s{ \rm min}}^2) = -0.42 \pm 0.02$.

In order to provide an estimate of the statistical significance of the above signal,  
we find it advantageous to recast our analysis in terms of 
the quantity $\Ls(r^2)$, thus capitalizing on the detailed error analysis 
applied to the lattice data of~\cite{Aguilar:2021lke}. Specifically, from the WI of \1eq{centeuc}  we will determine 
the form that $\Ls(r^2)$ would have if the null hypothesis were true,
and quantify its deviation from the actual lattice data.

Thus, 
setting $\Cfat(r^2) = 0$ into \1eq{centeuc}, we obtain the null hypothesis prediction for $\Ls(r^2)$, which we denote by $\Ln(r^2)$, given by 
\be 
\Ln(r^2) = F(0)\left\lbrace \frac{\w(r^2)\Delta^{-1}(r^2)}{r^2} + \widetilde{Z}_1 \left[ \Delta^{-1}(r^2)\right]^\prime \right\rbrace \,. \label{L0}
\ee
Substituting on the r.h.s. of \1eq{L0} the same ingredients as before, we obtain 
the $\Ln(r^2)$ shown as the green dotted line on the right panel of \fig{fig:Cres}.
The green band enveloping $\Ln(r^2)$ captures the error propagated from $\w(r^2)$;
it is obtained by using as inputs into \1eq{L0} the curves delimiting the 
blue band in the right panel of \fig{fig:intgnd_Ws}. 

The results shown in \fig{fig:Cres} demonstrate that the statistical error of the lattice 
cannot account for the discrepancy between $\Ls(r^2)$ and $\Ln(r^2)$; evidently, the null hypothesis is strongly
disfavored.

In order to quantify the above statement, we adopt the following procedure.

\begin{figure}[t]
\includegraphics[width=0.475\textwidth]{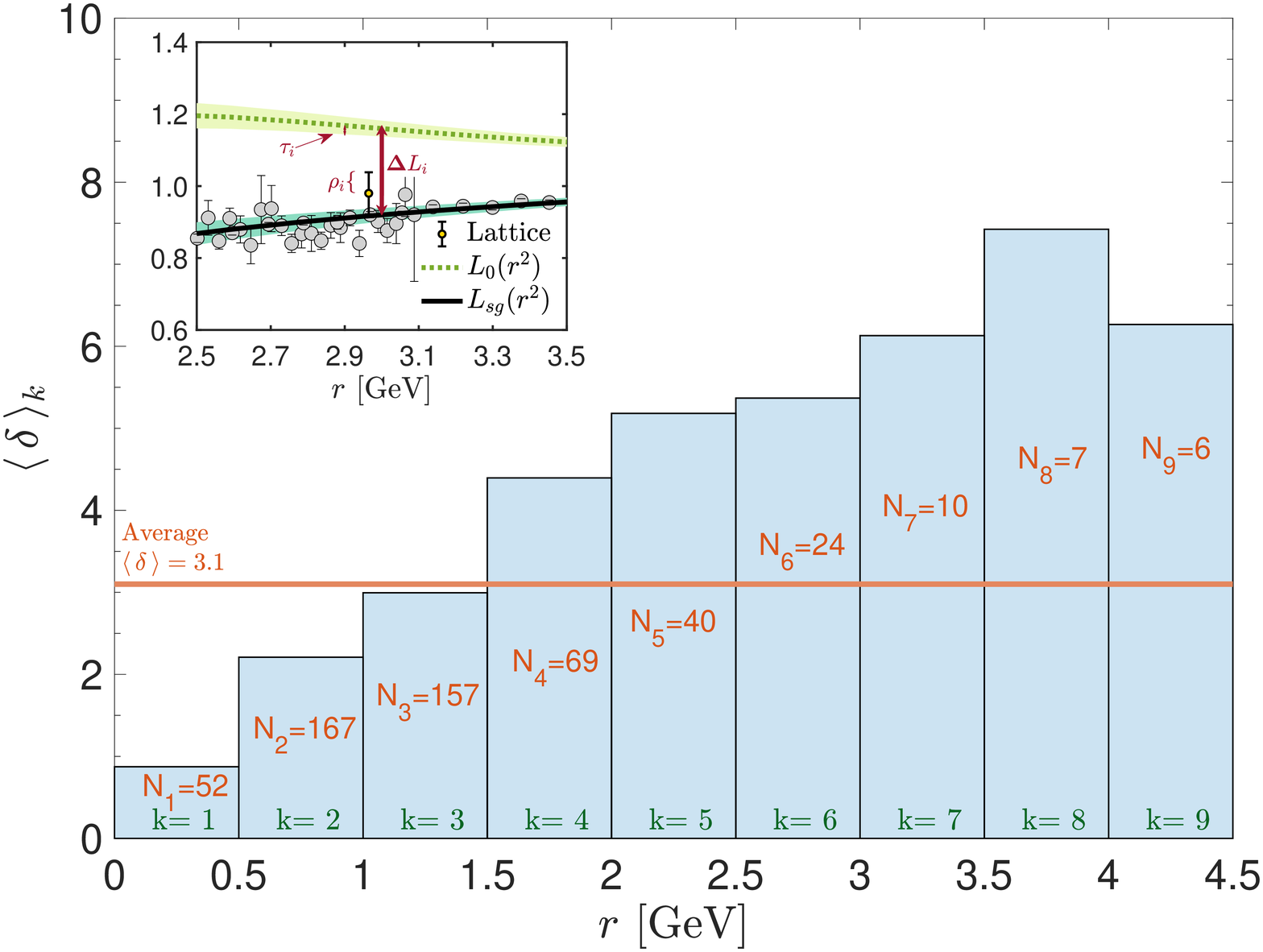}\hfill 	
\includegraphics[width=0.475\textwidth]{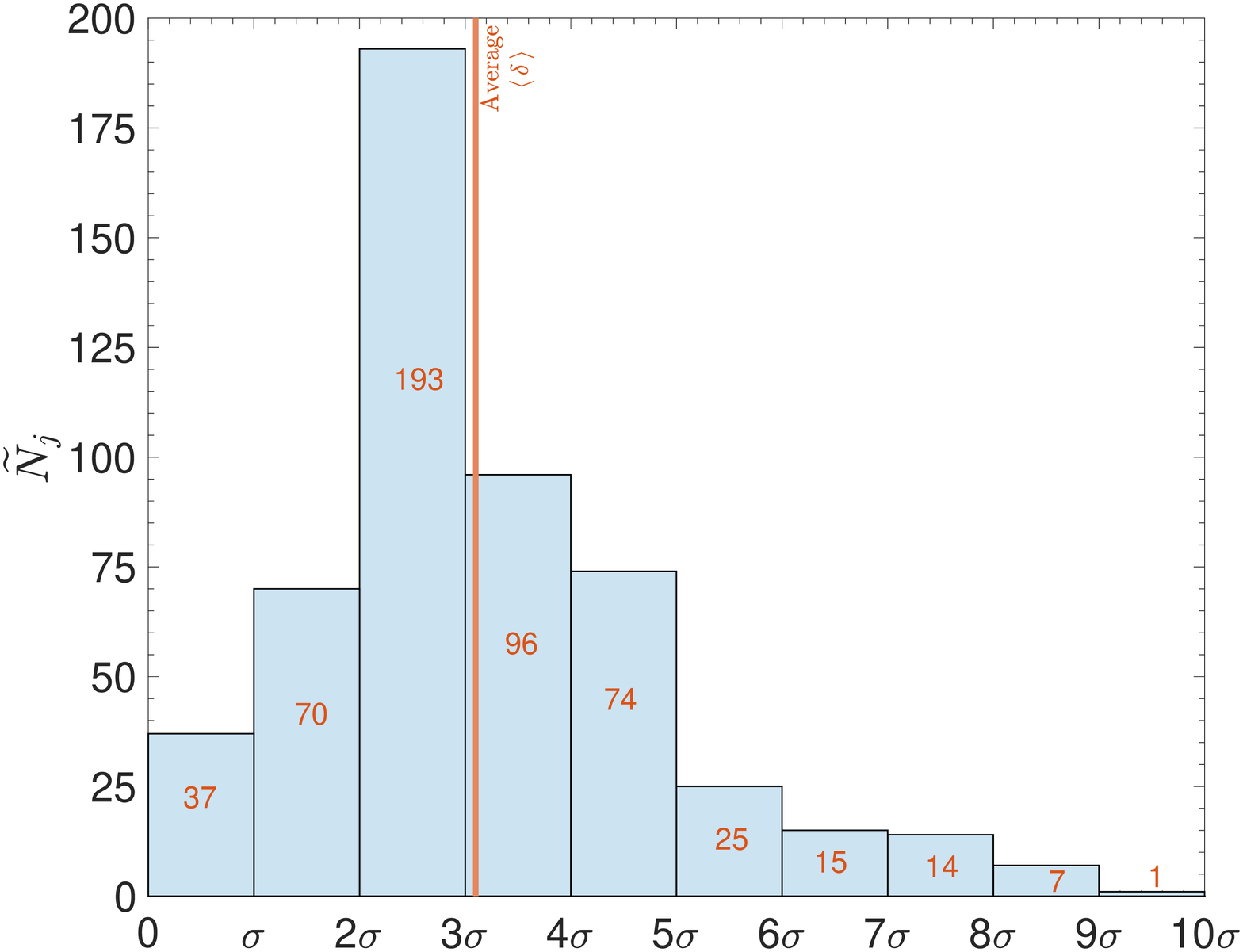}		
\caption{ Left: $\langle \,{\delta}\,\rangle_k$, the average standard deviation of all point belonging to the $k$-th bin of momentum,
  plotted as a function of the momentum $r$. The inset describes the definition
  of the local individual deviations $\rho_i$ and $\tau_i$, which are combined in a total correlated error \mbox{$\sigma_i = \rho_i + \tau_i$}, valid for any given point, as described in the text, items \mbox{({\it i}) - ({\it iii})}. 
Right:  The distribution  of all available points  into bins 
of a given standard deviation.  In both figures, the orange line represents the total average,  \mbox{$\langle\, \delta\,\rangle= 3.1\sigma$}.}
\label{fig:histo}
\end{figure}

({\it i}) At every data point, denoted by the index $i$ and located at the momentum $r_i$, we consider the standard error in the lattice data for $\Ls$, denoted by $\rho_i$, and the propagated error in the null hypothesis prediction, $\Ln(r_i)$, denoted by $\tau_i$, as shown in the inset of the left panel of \fig{fig:histo}.

({\it ii}) These errors are found to be \emph{correlated}. Specifically,  when using a higher $\Ls(r^2)$ as input in \1eq{L0},  we obtain a lower $\Ln(r^2)$.  Hence, the total error, denoted by $\sigma_i$, is given by $\sigma_i := \rho_i + \tau_i$. Note that the $\sigma_i$ so defined is larger than the corresponding errors combined in quadrature, \ie $\sqrt{ \rho_i^2 + \tau_i^2}$, which would be appropriate if the $\rho_i$ and $\tau_i$ were independent.

({\it iii}) Next, we measure the distance  
\mbox{$\Delta L_i := |\Ln(r_i^2) - \Ls(r_i^2)|$}, also shown in the inset of the left panel of \fig{fig:histo},
and then divide it by the corresponding total error, $\sigma_i$. 
The resulting ratio, \mbox{${\delta}_i := \Delta L_i/\sigma_i$}, measures the point-by-point deviation
between the two curves, computed in units of the $\sigma$ (standard deviation) assigned to every given data point.  

({\it iv}) 
The entire momentum range considered, \mbox{[$0, 4.5$] GeV}, is divided into nine equal
bins of length $0.5$ GeV; thus,
the $k$-th bin is defined as the interval \mbox{[$0.5\,(k-1), 0.5\, k$] GeV}, \mbox{$k=1,\cdots, 9$}.
In addition, we denote by $N_k$ the total number of points in the $k$-th bin; we
have \mbox{$N_k=(52,167,157,69,40,24,10,7,6)$,} accounting for a total of $N=532$ lattice points.

({\it v})
Then, we compute the average value of the ratio ${\delta}_i$ within the $k$-th bin, and denote the 
answer by \mbox{$\langle \,{\delta}\,\rangle_k$}, namely 
\be
\langle \,{\delta}\,\rangle_k = \frac{1}{N_k} \sum_{i=n_k + 1}^{n_k + N_k} \!\!{\delta}_i \,, \qquad \mathrm{ with} \,\,\,\, n_k = \sum_{\ell = 1}^{k - 1} N_\ell \,.
\label{Rk}
\ee

({\it vi})
Finally, the total average, $\langle\, \delta\,\rangle$, is defined as  
\be
\langle\, {\delta}\,\rangle = \frac{1}{N} \sum_{i=1}^{N} {\delta}_i\,,
\label{Rt}
\ee
and furnishes a measure of the global deviation between the signal [$\Ls(r^2)$] and the null hypothesis [$\Ln(r^2)$] curves. 

The outcome of this procedure is displayed in the left panel of \fig{fig:histo}, 
where the quantity $\langle \,{\delta}\,\rangle_k$, obtained at step ({\it v}), is plotted for each bin.
The value of $\langle\, {\delta}\,\rangle$, computed at step ({\it vi}), is  $\langle\, {\delta}\,\rangle = 3.1$,
and is marked by the orange horizontal line.

As we can observe, the values of  $\langle \,{\delta}\,\rangle_k$ for the bins with  $r > 1.5$~GeV are
considerably higher than $\langle\, \delta\,\rangle$. In fact, 
for $r > 2$~GeV the value of the corresponding $\langle \,{\delta}\,\rangle_k$ exceeds $5\sigma$; however, the available points in this interval are
relatively few.
The sizable signal found above $2$~GeV may be understood as follows. 
First, near $2$~GeV, the lattice curve $\Ls$ is the farthest away from its null hypothesis counterpart, $\Ln$,
leading to large values for the $\Delta L_i$ [see ({\it iii})]
in that region. Second, for $r > 3$~GeV, the $\Ls(r^2)$ and $\Ln(r^2)$
approach each other; nevertheless, since the lattice error bars become very small in the UV,
a rather strong signal emerges.

We next consider the distribution of all available points into bins of a given standard deviation, {\it regardless} of the momentum
assigned to each point. The length of each bin is one $\sigma$,
the $j$-th bin ($j=1,...10$) contains all points whose standard deviation lies in the interval $[j-1,\, j] \sigma$, 
and we denote the number of these points by ${\widetilde N}_{j}$.
The result of this grouping is shown in the right panel of \fig{fig:histo}.
We observe that, the largest number of points (193) is contained in the $[2-3] \sigma$  bin,  
while 133 points, corresponding to $25\%$ of the total number, are at or above the $5\sigma$ significance level.
The average of $3.1 \sigma$ is denoted by the orange vertical line.

Given that the truncation error in $\w(r^2)$ is the main uncertainty in our analysis, we end this section by
considering two interesting limiting cases associated with this function. 

First, given the clear proximity between $\Cfat(r^2)$ and $\CB(r^2)$, shown in the left panel of \fig{fig:Cres},
it is tempting to ask whether a small modification in the shape of $\w(r^2)$ could
make $\Cfat(r^2)$ and $\CB(r^2)$ agree perfectly.

To that end we substitute $\w(r^2)\to \wB(r^2)$ and $\Cfat(r^2)\to\CB(r^2)$ in \1eq{centeuc}
to obtain the function $\wB(r^2)$ necessary to reproduce $\CB(r^2)$. Specifically,
\be 
\wB(r^2) = r^2\Delta(r^2) \left\lbrace \frac{\Ls(r^2) - \CB(r^2)}{F(0)} - \widetilde{Z}_1\left[\Delta^{-1}(r^2)\right]^\prime \right\rbrace \,. 
\label{W_BSE}
\ee

\begin{figure}[t]
\includegraphics[width=0.475\textwidth]{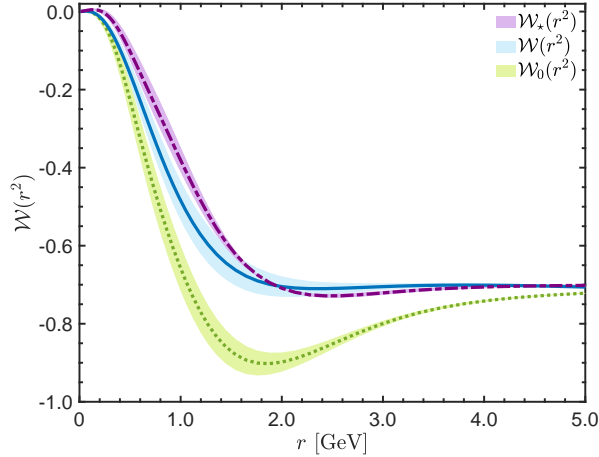} 		
\caption{ Direct comparison between the SDE result for $\w(r^2)$ (blue continuous)
  and the $\wB(r^2)$ (purple dot-dashed) that imposes the 
 equality between the $\Cfat(r^2)$ obtained from the WI and the BSE.
 Also shown is the $\wn(r^2)$ that corresponds to the null hypothesis (green dotted).
 The bands around each curve correspond to the propagated errors from the lattice $\Ls(r^2)$. }
\label{fig:W_tests}
\end{figure}

The $\wB(r^2)$ resulting from \1eq{W_BSE} is shown as the purple dot-dashed curve and the associated error band in \fig{fig:W_tests},
where it is compared to the SDE result for $\w(r^2)$. Indeed, we observe that a minor adjustment in the shape of
$\w(r^2)$ would bring $\Cfat(r^2)$ and $\CB(r^2)$ to a perfect agreement. 

Second, it is instructive to consider what would happen if the null hypothesis were valid,
and all resulting mismatches were to be absorbed exclusively into a modification of $\w(r^2)$, to be denoted by $\wn(r^2)$. 

Setting $\Cfat(r^2) = 0$ and $\w(r^2)\to\wn(r^2)$ into \1eq{centeuc}, we obtain
\be 
\wn(r^2) = r^2\Delta(r^2) \left\lbrace \frac{\Ls(r^2)}{F(0)} - \widetilde{Z}_1\left[\Delta^{-1}(r^2)\right]^\prime \right\rbrace \,. \label{W0}
\ee

In \fig{fig:W_tests} we show $\wn(r^2)$ as the green dotted curve. The band around it represents the propagated error of the lattice $\Ls(r^2)$;
it is obtained by substituting  $\Ls(r^2)$ in \1eq{W0} by the $\Ls^{\pm}(r^2)$ of \1eq{Lerr}. 

We note that the $\w(r^2)$ obtained from the SDE (blue solid curve)
is comfortably separated from $\wn(r^2)$. In fact, our attempts to obtain solutions of the SDE in the 
vicinity of $\wn(r^2)$ have been unavailing.

\section{\label{sec:conc} Discussion and Conclusions}

In the present work we have investigated in detail 
a characteristic feature that is intimately
linked with the onset of the Schwinger mechanism in QCD, and the ensuing 
emergence of an effective gluon mass scale.
The action of this mechanism relies on the
inclusion of massless longitudinal poles in the fundamental vertices of the theory,
which participate nontrivially in the realization of the corresponding STIs. 
This, in turn causes a distinct displacement to the WI satisfied by the pole-free part of the vertices involved,
quantified by the function $\Cfat(r^2)$,
which is formally identical to the bound-state wave function that governs the dynamical formation 
of the massless poles. We have computed $\Cfat(r^2)$ for the case of the three-gluon vertex
in two distinct ways: by solving the corresponding BSE and 
by appropriately combining the ingredients appearing in the non-Abelian WI.
In both cases we have relied predominantly on results obtained from lattice simulations,
with the exception of the special function $\w(r^2)$, which was determined from an SDE.  
The results found for  $\Cfat(r^2)$ are clearly nonvanishing and 
in excellent mutual agreement, providing additional support to the details of the general
dynamical picture put forth in a series of articles. In particular, the dual r$\hat{\textrm{o}}$le
played by  $\Cfat(r^2)$ is especially noteworthy, hinting towards deeper connections that have yet to be
unraveled.

It is important to stress that
the computation of the null hypothesis, presented in Sec.~\ref{sec:numan},  
proceeds by assuming that all inputs of \1eq{L0} 
retain their known form; in particular, salient features, 
such as the saturation of the gluon propagator and the ghost dressing
function, persist unaltered. In that sense, this specific implementation probes
the compatibility between the lattice results and the absence of a displacement in
the non-Abelian WI of the three-gluon vertex. Seen from this point of view, 
one might state that this particular possibility is excluded at the level of $3.1 \sigma$.

As mentioned in Appendix~\ref{sec:nbse}, the value of $\alpha_s$ that corresponds to the 
eigenvalue of the system is $\alpha_s=0.63$, which is considerably different from the
value $\alpha_s=0.27$ found within the ``asymmetric'' MOM scheme that we employ.
The discrepancy may be interpreted as a truncation artifact, given that the corresponding BSE kernels
has been approximated by their one-particle exchange diagrams, as depicted in \fig{fig:BSEs}.
In addition, the expressions employed for the fully dressed vertices comprising these kernels 
contain a certain amount of uncertainty. Quite interestingly, a preliminary 
numerical exploration indicates that minor modifications
of the kernel affect the value of $\alpha_s$ considerably, without practically modifying
the form of the solution found for $\CB(r^2)$ and $\Cc(r^2)$. This observation suggests that, while the
decrease of $\alpha_s$ towards its MOM value may require a more refined 
knowledge of the corresponding BSE kernels, the obtained solutions should be considered as fairly reliable.

The proximity between the $\w(r^2)$, computed in Appendix~\ref{sec:cw}, and the  $\wB(r^2)$
obtained from \1eq{W_BSE}, suggest that minor modifications of the
inputs used for the SDE of \1eq{Wsde_diags_Euc}
might lead to an even better coincidence. In this context, it is interesting to point out that
the determination of the transverse form factors $Y_i$ (see item ({\it iii})  in Appendix~\ref{sec:cw}) 
is subject to a considerable uncertainty, originating from the approximations implemented to the complicated 
SDE satisfied by  the three-gluon vertex (Fig.~6 in~\cite{Aguilar:2021lke}). Given the
relevance of $\w(r^2)$ for the
systematic scrutiny of the Schwinger mechanism, as exposed in the present work, it may be worthwhile 
revisiting this particular computation.

As mentioned below \1eq{stip}, the Schwinger mechanism induces poles also in the 
ghost-gluon kernel, $H_{\mu\nu}(r,q,p)$. This may be understood qualitatively by considering the 
diagrams ($d_1$) and ($d_2$) in \fig{fig:HSDE}: the fully-dressed ghost-gluon
and three-gluon vertices (with Lorentz index $\nu$ and incoming momentum $p$)
contain poles, which are transmitted to the form factors of $H_{\mu\nu}(r,q,p)$
associated with the tensorial structures $p_{\nu} p_{\mu}$ and $p_{\nu} r_{\mu}$.  
It would be important to compute in detail the pole structure of $H_{\mu\nu}(r,q,p)$, especially in view of the 
STI \mbox{$r^\mu H_{\mu\nu}(r,q,p) = \Gamma_\nu(p,r,q)$}, which links nontrivially
the form factors of $H_{\mu\nu}(p,r,q)$ and $\Gamma_\mu(p,r,q)$, in general, and the corresponding pole terms, 
in particular. Specifically, one may explore how accurately the appropriate
combination of pole terms coming from $H_{\mu\nu}(p,r,q)$ will reproduce the corresponding term contained in 
$\Gamma_\mu(p,r,q)$. We hope to undertake such a study in the near future.

The scale ambiguity associated with the BSE amplitudes $\Cfat(r^2)$ and $\C(r^2)$ results in from considering only the leading order terms of the BSEs in an expansion around $q = 0$, which furnishes homogeneous linear equations. In general kinematics, however, the presence of inhomogeneous terms in the BSEs
resolves this ambiguity. As such, the scales of $\Cfat(r^2)$ and $\C(r^2)$ can be fixed by taking the \mbox{$q = 0$} limit of the solution of the corresponding inhomogeneous BSEs, treated beyond leading order in $q$. In the context of conventional bound states, this procedure is well understood~\cite{Nakanishi:1969ph} and explicit scale-setting equations have been derived, which are sometimes referred to as ``canonical normalization condition''~\mbox{\cite{Maris:1997tm,Blank:2010bp}}. It is our intention to pursue this point in an upcoming study, and settle dynamically
the scale of the corresponding solutions.

\section*{Acknowledgments}
\label{sec:acknowledgments}
The work of  A.~C.~A. is supported by the Brazilian CNPq grants 307854/2019-1 and 464898/2014-5 (INCT-FNA).
A.~C.~A. and M.~N.~F.  also acknowledge financial support from  the FAPESP projects 2017/05685-2 and 2020/12795-1, respectively.   J.~P. is supported by the  Spanish AEI-MICINN grant PID2020-113334GB-I00/AEI/10.13039/501100011033,
and the  grant  Prometeo/2019/087 of the Generalitat Valenciana.

\appendix

\section{\label{sec:nbse} Technical details on the BSE system}

  
In this Appendix we present details related to the numerical treatment of the BSE system
formed by $\Cfat(r^2)$ and $\C(r^2)$, shown in \fig{fig:BSEs}. 

For the Bose symmetric three-gluon vertex appearing in the diagrams of \fig{fig:BSEs} 
we employ the tensor basis of Ball-Chiu~\cite{Ball:1980ax,Aguilar:2019jsj},
\be 
\g^{\alpha\mu\nu}(q,r,p) = \sum_{i = 1}^{10} X_i(q,r,p) \ell_i^{\alpha\mu\nu} + \sum_{i = 1}^{4} Y_i(q,r,p) t_i^{\alpha\mu\nu} \,, \label{BC_basis}
\ee
where the explicit form of the basis tensors
$\ell_i^{\alpha\mu\nu}$ and $t_i^{\alpha\mu\nu}$ is given in Eqs.~(3.4) and (3.6) of~\cite{Aguilar:2019jsj}. At tree level, $X_1^{(0)} = X_4^{(0)} = X_7^{(0)} = 1$, while all other $X_i^{(0)}$ and $Y_i^{(0)}$ vanish.

It is convenient to introduce 
the transversely projected vertex, $\overline{\g}_{\alpha\mu\nu}(q,r,p)$, defined as 
\be 
\overline{\g}_{\,\alpha\mu\nu}(q,r,p) := P^{\alpha'}_\alpha(q)P^{\mu'}_\mu(r)P^{\nu'}_\nu(p) \g_{\alpha'\mu'\nu'}(q,r,p) \,. \label{gbar_def}
\ee
Similarly, the tensorial decomposition of the vertex $\Gamma_\mu(q,r,p)$ is given by 
\be 
\Gamma_\mu(q,r,p) = r_\mu B_1(r,p,q) + q_\mu B_2(r,p,q) \,;
\label{Acc_Basis}
\ee
at tree level, $B_1^{(0)} = 1$ and $B_2^{(0)} = 0$. The ghost-anti-ghost symmetry of $\Gamma_\mu(q,r,p)$ in the Landau gauge implies that $B_1(r,p,q) = B_1(p,r,q)$.


Next, we pass to Euclidean space and employ spherical coordinates. For convenience, we define the variables $x := r^2$, $y := k^2$, and \mbox{$u := (r-k)^2 = x + y - 2\sqrt{xy}c_\phi$}, with $\phi$ denoting the angle between the momenta $k$ and $r$, while $s_\phi := \sin \phi$, $c_\phi := \cos \phi$. Furthermore, we parametrize scalar form factors, such as $B_1(q,r,p)$, in terms of the squares of their first two arguments and the angle between them, \eg \mbox{$B_1(-r,k,r-k)\to B_1(x,y,\pi - \phi)$}.
Then, the final set of BSEs reads
\bea 
\Cfat(x) &=& \lambda \left[ \int_0^\infty \!\!\!dy \, y \sqrt{x y} \Delta^2(y) \Cfat(y) \int_0^\pi  \!\!\!d\phi s_\phi^2 c_\phi \Delta(u) {\cal N}(x,y,u) 
\right. \nonumber\\ &&
 \left. \hspace{2cm}- 2 \int_0^\infty  \!\!\!dy \sqrt{\frac{y}{x}}F^2(y) \C(y) \int_0^\pi  \!\!\!d\phi s_\phi^4 c_\phi \frac{F(u)}{u} B_1^2(u,y,\chi) \right] \,, \nonumber\\
\C(x) &=& 3\lambda \left[ \int_0^\infty\!\!\!dy \, y \sqrt{xy} \Delta^2(y) \Cfat(y)  \int_0^\pi \!\!\! d\phi s_\phi^4 c_\phi \frac{F(u)}{u} B_1^2(x,u,\theta) \right. \nonumber\\&& 
\left.\hspace{2cm} + \int_0^\infty\!\!\!dy \sqrt{xy} F^2(y) \C(y) \int_0^\pi  \!\!\!d\phi s^4_\phi c_\phi \frac{\Delta(u)}{u} B_1^2(x,y,\pi - \phi) \right] \,. 
\label{CsBSE_Euc}
\eea
In the above equation, 
\be 
\lambda := \frac{\alpha_s C_{\rm A}}{12\pi^2} \,,
\label{lambda_def}
\ee
where $\alpha_s := g^2/4\pi$.
The angles $\chi$ and $\theta$ are given by
\be 
\chi = \cos^{-1}\left( \frac{\sqrt{x}\cos\phi - \sqrt{y}}{\sqrt{u} } \right) \,, \qquad \theta = \cos^{-1}\left( \frac{\sqrt{y}\cos\phi - \sqrt{x}}{\sqrt{u} } \right) \,. 
\label{angs}
\ee
Finally, 
\bea 
{\cal N}(x,y,u) := -\frac{1}{x}\left[ \overline{\g}_{\,\alpha\beta\gamma}(-r,k,r-k)\overline{\g}^{\,\alpha\beta\gamma}(-r,k,r-k) \right]_{\rm E} \,, 
\label{NKern}
\eea
where the subscript ``E'' indicates that \1eq{NKern} is to be converted to Euclidean coordinates.

The ${\cal N}(x,y,u)$ of \1eq{NKern} can be written in terms of the form factors $X_i$ and $Y_i$ of \1eq{BC_basis}; note that 
the $X_i$ with $i = 2,5,8,10$ drop out, because they are annihilated by the transverse projection in \1eq{gbar_def}.

  Then, we obtain
\bea
{\cal N}(x,y,u) &=& \frac{s^2_\phi}{ 2 u^2 x^2 y } \left\lbrace u T_1 \left[ T_1 \left( ( u - x - y )^2 + 2xy \right) + 4 x y ( T_2 + T_3 ) + 4 T_4 ( u - x - y ) \right] \right. \nonumber\\
&& + x T_2 \left[ T_2 \left( ( u - x + y )^2 + 2 u y ) \right) + 4 u y T_3 - 4 T_4 ( u - x + y ) \right] \\
&& \left. + y T_3 \left[ T_3 \left( ( u + x - y )^2 + 2 u x \right) - 4 T_4 ( u + x - y ) \right] + 4 T_4^2 ( u + x + y ) \right\rbrace \,, \nonumber
\eea
where 
\bea 
T_1 &=&  - u( X_1 - X_4 - X_7 + x y Y_1 ) - x( X_1 - X_4 + X_7  - 2 y X_3 ) - y( X_1 + X_4 - X_7 ) \,, \nonumber \\
T_2 &=& u( X_1 - X_4 - X_7 + 2 y X_6 - x y Y_2 ) + x ( X_1 - X_4 + X_7 ) - y ( X_1 + X_4 - X_7 ) \,, \nonumber \\
T_3 &=& u( X_1 - X_4 - X_7 + 2 x X_9 - x y Y_3 ) - x ( X_1 - X_4 + X_7 ) + y( X_1 + X_4 - X_7 ) \,, \nonumber \\
T_4 &=& \frac{1}{2}\left\lbrace u^2( X_1 - X_4 - X_7 ) - ( x - y )\left[ x ( X_1 - X_4 + X_7 ) - y( X_1 + X_4 - X_7 ) \right]\right. \nonumber\\
&& \left. + 2 u \left[ x ( X_7 - y Y_4 ) + y X_4 \right] \right\rbrace \,,
\eea
and we suppress the functional dependence \mbox{$X_i \equiv X_i( x, y, \pi - \phi )$} and \mbox{$Y_j \equiv Y_j( x, y, \pi - \phi )$}.

For the numerical evaluation of \1eq{CsBSE_Euc}, in addition to the lattice propagators given in Appendix~\ref{sec:inp}, we need 
$B_1$, together with all $X_i$ and $Y_i$ that comprise ${\cal N}(x,y,u)$, in general kinematics.
These form factors are obtained as follows: 

({\it i}) For the $B_1$ we employ recent results (see Fig. 6  of~\cite{Aguilar:2021okw}),
obtained from an SDE analysis that uses as inputs 
lattice data that have been cured from volume and discretization artifacts. 

\begin{figure}[t]
\includegraphics[width=0.475\textwidth]{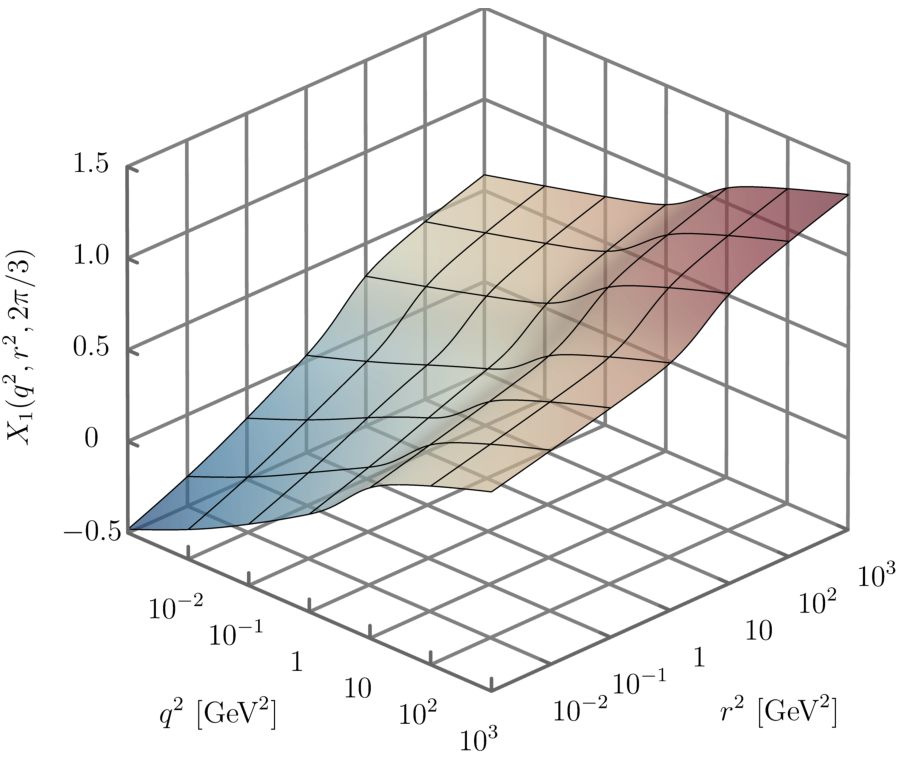}\hfill
\includegraphics[width=0.475\textwidth]{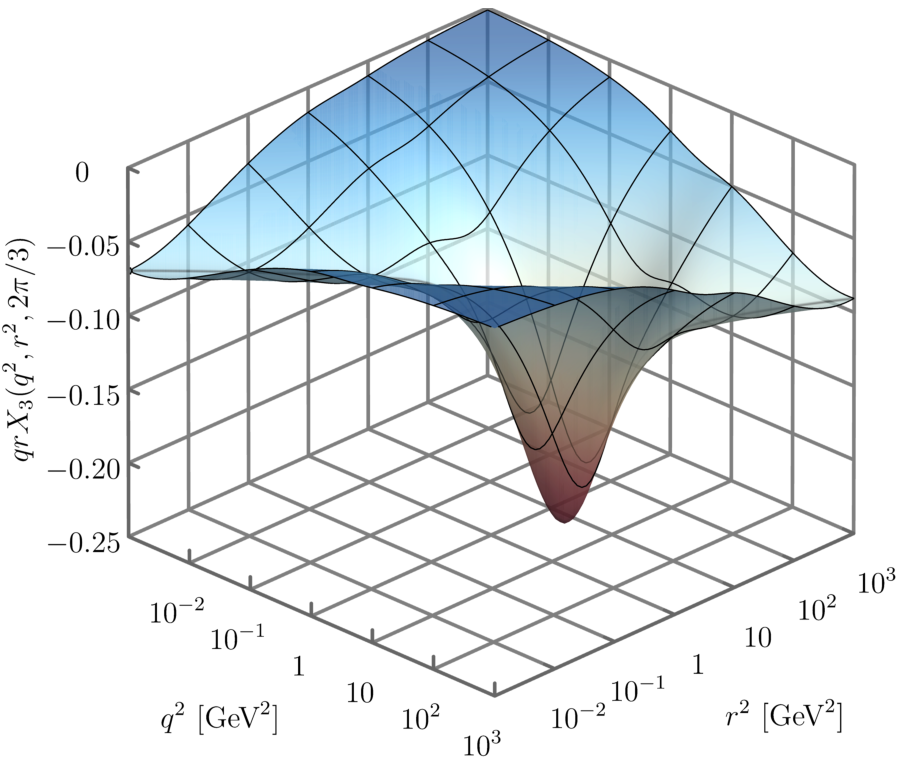} \\
\includegraphics[width=0.475\textwidth]{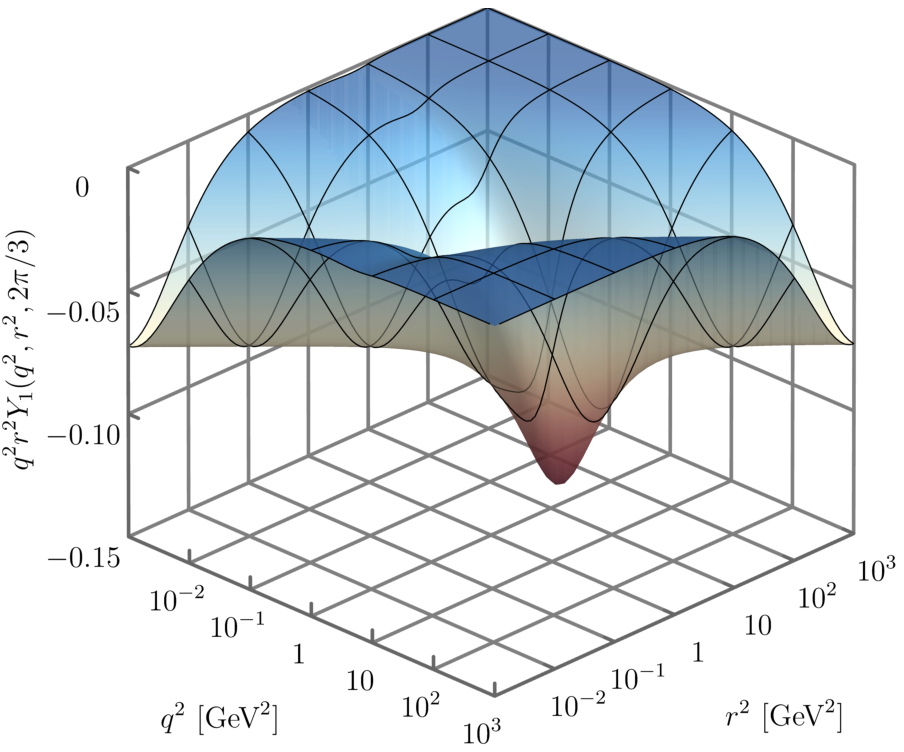}\hfill
\includegraphics[width=0.475\textwidth]{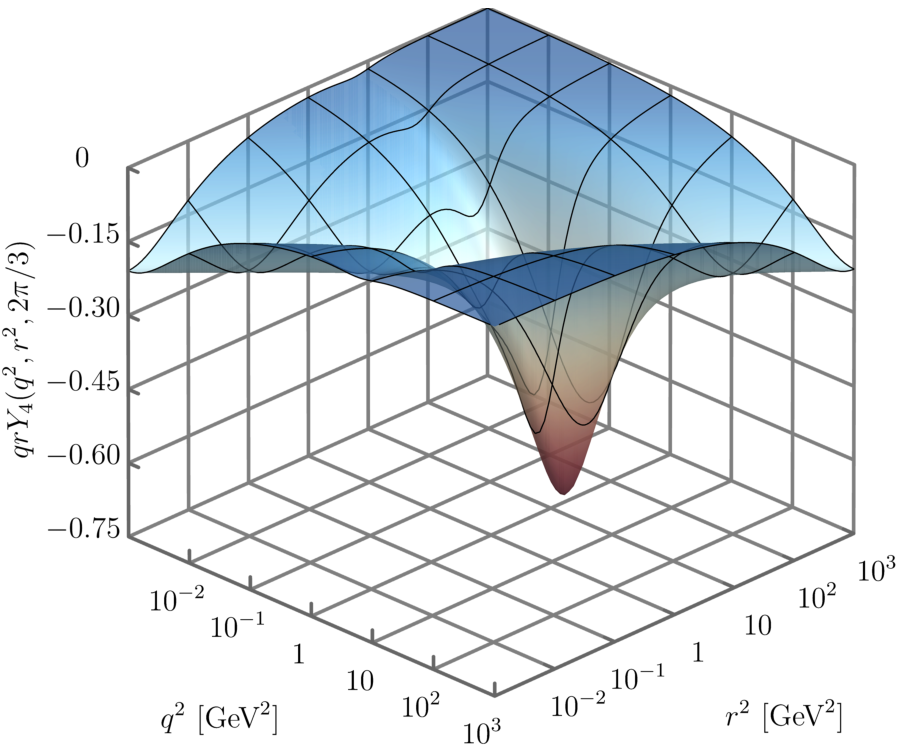} 		
\caption{The three-gluon vertex form factors used in the BSE system:  
$X_1(q^2,r^2,2\pi/3)$ (top left), $qrX_3(q^2,r^2,2\pi/3)$ (top right), $q^2r^2Y_1(q^2,r^2,2\pi/3)$ (bottom left), and  $qrY_4(q^2,r^2,2\pi/3)$ (bottom right).}
\label{fig:Xi_3D}
\end{figure}

({\it ii})
The $X_i$ are obtained from the nonperturbative generalization of the Ball-Chiu solution; the relevant formulas are given
in Eq.~(3.11) of~\cite{Aguilar:2019jsj}, and involve the ghost dressing function, the kinetic term of the gluon propagator,
and certain components of the ghost-gluon kernel.
Note that the inputs have been calibrated to exactly reproduce the lattice projection $\Ls(r^2)$, through the relation 
\be 
\Ls(r^2) = X_1(r^2,r^2,\pi) - r^2 X_3(r^2,r^2,\pi) \,.
\label{L_Xi} 
\ee
In this indirect way, the error bars assigned to the lattice calculation of $\Ls(r^2)$, encompassed by the functions $\Ls^{\pm}(r^2)$ of \1eq{Lerr},
find their way into our BSE determination of $\Cfat(r^2)$ and $\C(r^2)$, giving rise to the errors band indicated in Fig. \ref{fig:Cres}

({\it iii}) 
For the transverse components $Y_i$, which cannot be deduced from the fundamental STIs,
we resort to a SDE determination, along the lines of the analysis presented in~\cite{Aguilar:2021lke}; see, in particular, Fig.~6 therein.

({\it iv}) The results for 
$X_1(q^2,r^2,\theta)$, $qrX_3(q^2,r^2,\theta)$, $q^2r^2Y_1(q^2,r^2,\theta)$, and $qrY_4(q^2,r^2,\theta)$
are shown in \fig{fig:Xi_3D}, for the special case $\theta = 2\pi/3$;
$q$ and $r$ denote now the magnitudes of the corresponding Euclidean momenta.

({\it v})
By virtue of the Bose symmetry of the three-gluon vertex, 
the remaining form factors of the three-gluon vertex entering in \1eq{NKern} can be obtained from those shown in \fig{fig:Xi_3D} by appropriate permutations of their arguments, as explained in~\cite{Aguilar:2019jsj}.

Employing the ingredients described above, 
we solve the coupled system of BSEs of \1eq{CsBSE_Euc} numerically, 
obtaining the $\Cfat(r^2)$ and $\C(r^2)$ shown in \fig{fig:C_gl_Cgh}, together with the corresponding error estimates.

Since \1eq{CsBSE_Euc} does not have an inhomogeneous term 
and is linear in $\Cfat(r^2)$ and $\C(r^2)$, it corresponds to an eigenvalue problem. The resulting 
eigenvalues correspond to \mbox{$\alpha_s = 0.63 \mp 0.05$}, with signs opposite to those of the $\Ls^\pm(r^2)$ error bands, \ie using $X_i$ corresponding to a higher $\Ls(r^2)$ leads to a smaller $\alpha_s$. 
In these results, the overall constant was determined by matching the BSE prediction for $\Cfat(r^2)$ to the result obtained from the WI,
as explained in Sec.~\ref{sec:numan}.

\section{\label{sec:cw}Computation of $\w(r^2)$}

The function  $\w(r^2)$ is a central ingredient for our analysis, whose determination proceeds
through the study of the corresponding SDE.
In this Appendix we present the technical details related to this calculation, and discuss the validity of our
approximations.

\subsection{\label{sec:cw1} SDE, inputs, and solution}

\begin{figure}[t]
\includegraphics[width=1\textwidth]{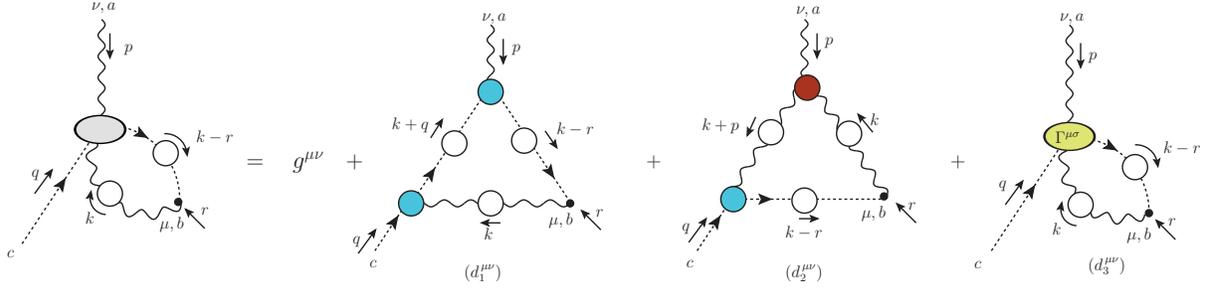}	
\caption{ The SDE satisfied by the ghost-gluon scattering kernel, $H_{\mu\nu}(r,q,p)$. }
\label{fig:HSDE}
\end{figure}

The starting point of our determination of $\w(r^2)$ is the SDE for the ghost-gluon scattering kernel, $H_{\mu\nu}(r,q,p)$, shown in \fig{fig:HSDE}, which is truncated at the one-loop dressed level, retaining only diagrams $(d_1)$ and $(d_2)$.
In the Landau gauge it is immediate to factor out of these two diagrams the ghost momentum $q$, in order to 
obtain $K_{\mu\nu\alpha}(r,q,p)$, in accordance with \1eq{HtoK}.
Finally, recalling \1eq{HKtens}, $\w(r^2)$ is obtained by isolating the $g_{\mu\nu}r_\alpha$ form factor of $K_{\mu\nu\alpha}(r,0,-r)$ and using \1eq{KW}.

Since a detailed derivation and renormalization of the $\w(r^2)$ equation has been carried out in Sec.~6 of Ref.~\cite{Aguilar:2020uqw}, here we only collect the main results. Specifically, we obtain the general expression
\be 
\w(r^2) = \w_1(r^2) + \w_2(r^2) \,,
\label{Wsde_Mink}
\ee
where $\w_1$ and $\w_2$ denote the contributions from diagrams $(d_1)$ and $(d_2)$ of \fig{fig:HSDE}, respectively, given by 
\bea 
\w_1(r^2) &=&  - \frac{ i g^2 C_{\rm A}\widetilde{Z}_1}{6} \int_k \Delta(k^2) D(k^2) D(t^2) B_1( t, k, - r)B_1( k, 0, - k) (r\cdot k)f(r,k) \,, \nonumber \\
\w_2(r^2) &=& - \frac{ i g^2 C_{\rm A}\widetilde{Z}_1}{6} \int_k \Delta(k^2) \Delta(t^2) D(t^2) B_1( t, 0, -t ) r^\alpha \overline{\g}^{\,\mu}_{\mu\alpha}(-r,k, t)\,, 
\label{Wsde_diags_Mink}
\eea
where 
\be
 f(r,k):= 1 -\frac{(k\cdot r)^2}{k^2r^2}\,,
\label{fkr}
\ee
and $\overline{\g}$ is defined in \1eq{gbar_def}.  Note the appearance in \1eq{Wsde_diags_Euc} of $\widetilde{Z}_1$, defined below \1eq{HtoK}, which
implements the renormalization of the SDE in the asymmetric MOM scheme~\cite{Aguilar:2020uqw}.

To express \1eq{Wsde_Mink} in Euclidean space, we use spherical coordinates and the kinematic variables $x$, $y$ and $u$ defined above \1eq{CsBSE_Euc}. Then, using the Ball-Chiu tensor basis of \1eq{BC_basis} for the three-gluon vertex, we obtain
\be 
\w(x) = \w_1(x) + \w_2(x) \,, \label{Wsde}
\ee
with
\bea 
\w_1(x) &=& \lambda\widetilde{Z}_1 \int_0^\infty \!\!\!\!\! dy \sqrt{xy} \Delta(y) F(y) B_1( y, 0, 0 ) \int_0^\pi \!\!\!d\phi \, c_\phi s_\phi^4 \frac{F(u)}{u} B_1( u, y, \chi) \,, \nonumber \\
\w_2(x) &=& - 2\lambda\widetilde{Z}_1 \int_0^\infty\!\!\!\!\! dy \, y\sqrt{xy} \Delta(y) \int_0^\pi \!\!\!d\phi s_\phi^4 \Delta(u) B_1( u, 0, 0 ) \frac{F(u)}{u^2} {\cal K}(x, y, u)\,, 
\label{Wsde_diags_Euc}
\eea
where $\lambda$  is defined in \1eq{lambda_def},   $\chi$ in \1eq{angs}, and we define the kernel ${\mathcal K}(x,y,u)$ as
\bea
{\mathcal K}(x,y,u) &:= &  \sqrt{x\,y}( c_\phi^2 + 2 ) X_1 
 + ( y - c_\phi \sqrt{x\,y} )c_\phi X_4 + ( x - c_\phi \sqrt{x\,y} )c_\phi X_7 - 3 c_\phi x \, y X_3   \nonumber \\
&&  - c_\phi y \, u X_6 - c_\phi x \, u X_9  + \frac{1}{2}{c_\phi x \, y \, u}( 3 Y_1 + Y_2 + Y_3 ) - u \sqrt{x \,y}\, Y_4  \,,
\label{fullK}
\eea
 where, again, \mbox{$X_i \equiv X_i( x, y, \pi - \phi )$} and \mbox{$Y_j \equiv Y_j( x, y, \pi - \phi )$}. 

Using the Bose symmetry relations involving permutations of arguments of the $X_i$ and $Y_i$, given by Eqs.~(3.7) to (3.10) of \cite{Aguilar:2019jsj}, it is possible to show that $\mathcal{K}(x,y,u)$ is symmetric under the exchange of $x\leftrightarrow y$.

Then, for the gluon propagator $\Delta(r^2)$ and the ghost dressing function $F(r^2)$ we use the fits presented in Appendix~\ref{sec:inp},  
while for the vertex form factors, $B_1$, $X_i$ and $Y_i$ we use the same inputs employed for the solution of the BSE of \1eq{CsBSE_Euc}.
Note that, as mentioned in Appendix~\ref{sec:inp}, 
all inputs are renormalized within the ``asymmetric'' MOM scheme, at the renormalization point \mbox{$\mu =4.3$ GeV}, for which  \mbox{$\alpha_s = 0.27$}.

\begin{figure}[t]
 	\includegraphics[width=0.475\textwidth]{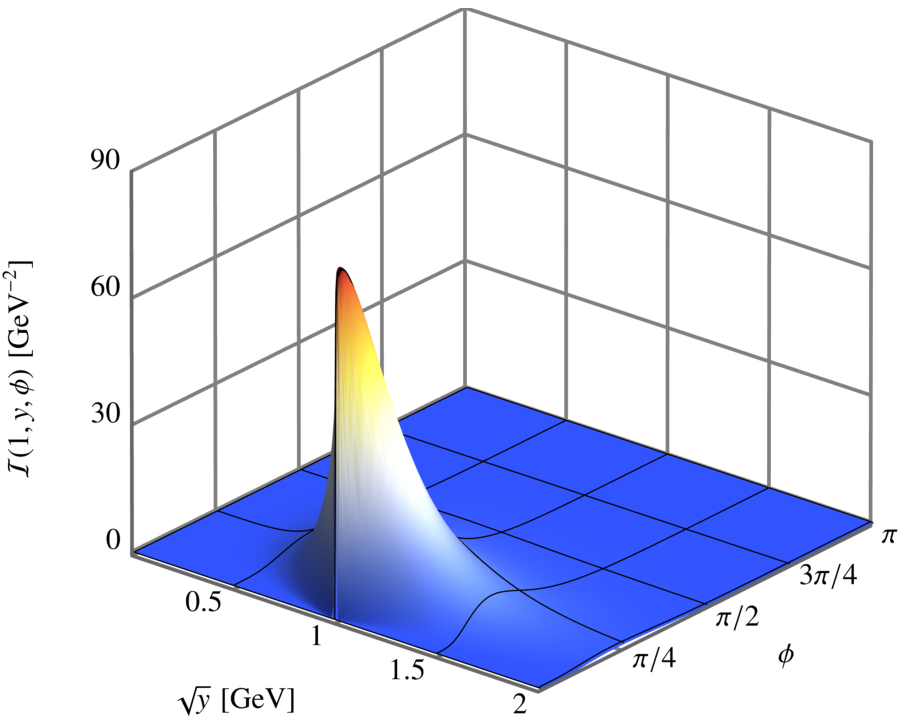}\hfill
 	\includegraphics[width=0.475\textwidth]{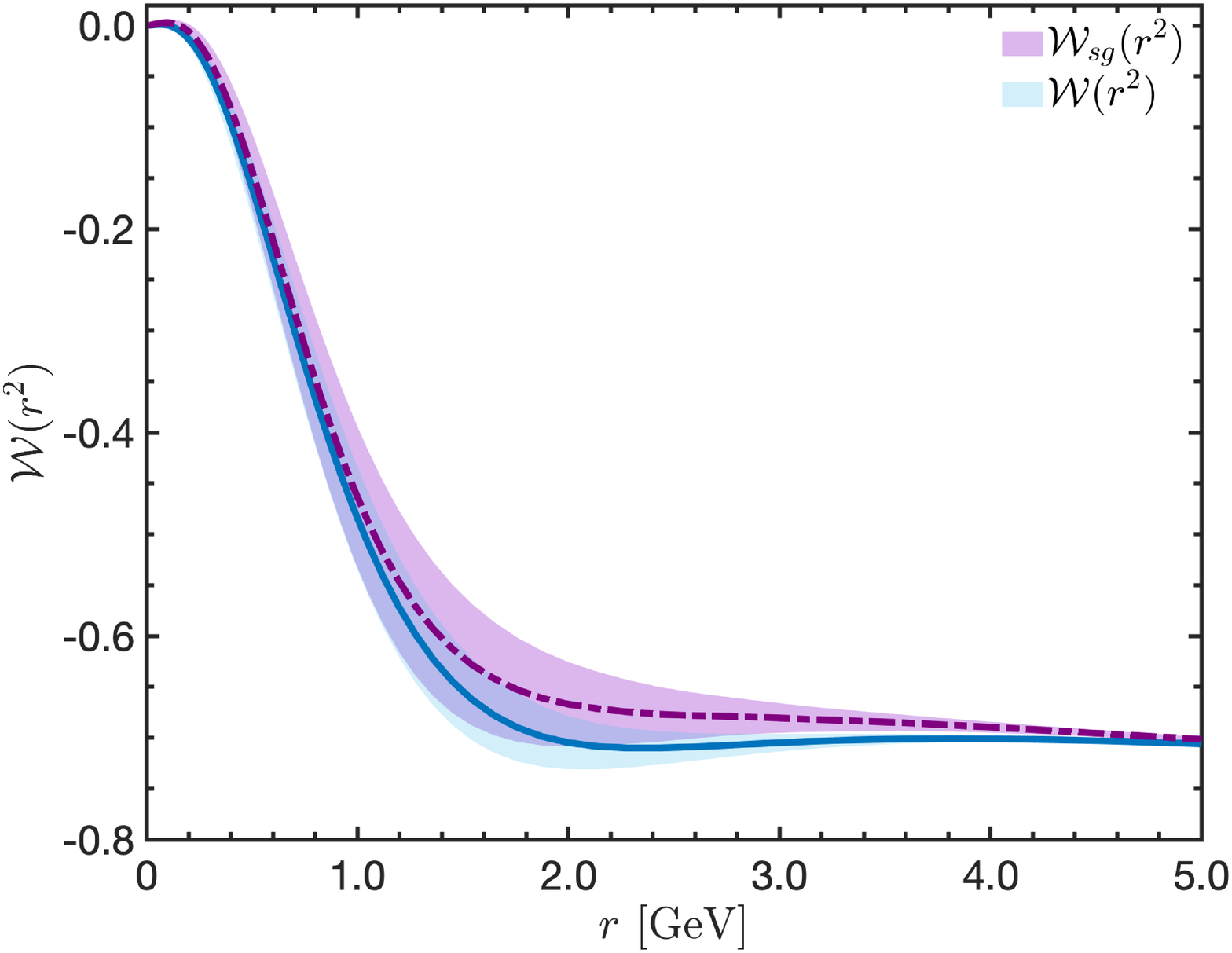}		
 	\caption{ Left: The integrand of $\w_2(x)$, $\Ical(x,y,\phi)$, given by \1eq{intgnd}, for $x = 1$~GeV$^2$. Note that its maximum occurs when $y$ is first set to $y = x = 1$~GeV and then $\phi = 0$, corresponding to $u = 0$. $\Ical(x,y,\phi)$ is intensely peaked around the maximum,  
dropping rapidly to zero away from it. Right: Comparison of $\w(r^2)$, obtained with \1eq{Wsde} (blue solid line), to $\ws(r^2)$, obtained with the replacement \eqref{K_s} into \1eq{Wsde} (purple dot-dashed line). The bands correspond to propagated error from the lattice $\Ls(r^2)$ of \cite{Aguilar:2021lke}. } 
\label{fig:intgnd_Ws}
\end{figure}

With these ingredients, we obtain for $\w(r^2)$ the blue continuous line shown in the right panel of \fig{fig:intgnd_Ws}. The blue band around it corresponds to error propagated from the uncertainty in the lattice $\Ls(r^2)$, through the same procedure explained in item ({\it ii}) of Section~\ref{sec:nbse}.

Given that $\w(r^2)$ is one of the main ingredients in the analysis of \1eq{centeuc}, it is important to consider in more detail the uncertainty in our SDE determination of this function.
It turns out that the contribution $\w_1(r^2)$ of \1eq{Wsde_diags_Euc} is negligible in comparison to $\w_2(r^2)$ (see Fig.~7 of \cite{Aguilar:2020uqw}), except for $r<0.5$~GeV, where $\w_2(r^2)$ decreases significantly.
Furthermore, 
diagram $(d_3)$ of \fig{fig:HSDE}, with the four-particle correlation function $\Gamma_{\mu\sigma}$ nested in it, is known to affect the 
ghost-gluon vertex only by $2\%$~\cite{Huber:2017txg}; thus, its omission is expected to have an insignificant effect 
on $\w(r^2)$. Therefore, the main uncertainty originates from the term 
$\w_2(r^2)$ of \1eq{Wsde_diags_Euc}, and is related to our incomplete knowledge of the 
form factors $X_i$ and $Y_i$ for general kinematics. 

In this regard, an examination of the integrand of $\w_2(r^2)$ in \1eq{Wsde_diags_Euc} shows that this contribution is dominated by the projection $\Ls(r^2)$ of the full three-gluon vertex that it contains. In turn, this observation suggests 
that the SDE determination of $\w(r^2)$ should be fairly accurate provided that the \emph{Ansatz} employed for the general kinematics three-gluon vertex reproduces in the soft-gluon limit the $\Ls(r^2)$ obtained on the lattice~\cite{Aguilar:2021lke}.

\subsection{\label{sec:cw2} A closer look at the SDE kernel}

To elucidate this last point, denote by $\Ical(x,y,\phi)$ the integrand of $\w_2(x)$ in \1eq{Wsde_diags_Euc}, \ie
\be 
\Ical(x,y,\phi) := s_\phi^4 y\sqrt{xy} \Delta(y) \Delta(u) \frac{F(u)}{u^2} B_1( u, 0, 0 ) {\cal K}(x, y, u) \,. \label{intgnd}
\ee
Then, since $F(u)$ and, especially, $\Delta(u)$ are decreasing functions of $u$, the term $\Delta(u)F(u)/u^2$ in the second line of \1eq{intgnd} causes the $\Ical(x,y,\phi)$ to decrease rapidly at large $u$. Hence, $\w_2(x)$ should be dominated by the small $u$ region of its integrand.

Next, recalling that \mbox{$u = x + y - 2\sqrt{xy}c_\phi$}, we note that $u = 0$ occurs when $y = x$ and $\phi = 0$ simultaneously.
Also, we emphasize that, in spite of the presence of the factor $u^2$ in the denominator, \1eq{intgnd} is finite at $u = 0$, due to the vanishing of $s^4_\phi$ when $\phi = 0$.

In the left panel of \fig{fig:intgnd_Ws} we plot $\Ical(x,y,\phi)$ for $x = 1$~GeV$^2$ and general $y$ and $\phi$, using the general kinematics $X_i$ and $Y_i$ of \fig{fig:Xi_3D} in the evaluation of ${\cal K}(x,y,u)$. There we confirm that $\Ical(x,y,\phi)$ is largest around $y = x = 1$~GeV$^2$ and $\phi = 0$, decaying rapidly to zero at large $u$. Other values of $x$ lead to similar surfaces, with pronounced peaks at $y = x$.

Due to the sharply peaked structure of $\Ical(x,y,u)$, one expects that the value of the integral defining $\w_2(x)$ in \1eq{Wsde_diags_Euc}
should depend mainly on the maximum value of $\Ical(x,y,u)$. To determine the value of this maximum, we expand \1eq{intgnd} around $y = x$, and finally around $\phi = 0$.\footnote{The limit 
  of $\Ical(x,y,\phi)$ as $u \to 0$ is path-dependent, vanishing if $\phi = 0$ is set first; however, we are interested in its maximum, which occurs when $y = x$ is set first, and $\phi = 0$ after, as is clear from \fig{fig:intgnd_Ws}.} To this end, first note that
\be 
  \lim_{\substack{\phi \to 0 \\ y\to x}} \left( \frac{s_\phi^4}{u^2} \right)= \frac{1}{x^2} \,.
\label{lim}
\ee
The limit of the other terms in \1eq{intgnd} as $u \to 0$ is straightforward, leading to
\be 
  \lim_{\substack{\phi \to 0 \\ y\to x}}   \Ical(x,y,u)  = 3 \Delta(0) F(0) B_1(0,0,0) x \Delta(x) \left[ X_1(x,x,\pi) - x X_3(x,x,\pi) \right] \,, \label{intgnd_u0_first}
\ee
which, after using \1eq{L_Xi} becomes 
\be 
 \lim_{\substack{\phi \to 0 \\ y\to x}}  \Ical(x,y,u)  = 3 \Delta(0) F(0) B_1(0,0,0) x \Delta(x) \Ls(x) \,. \label{intgnd_u0}
\ee
Given that the 
$X_i$ employed reproduce $\Ls(x)$ when the combination in \1eq{L_Xi} is formed, 
the above considerations indicate that our result for $\w(r^2)$ should be rather accurate. 
Essentially $\w_2(x)$ appears dominated by the ``slice'' that corresponds to $\Ls(r^2)$, with little
or no effect from all other kinematic configurations.

\subsection{\label{sec:cw3} A special Ansatz for the three-gluon vertex}

To confirm this hypothesis explicitly, 
we compute $\w(r^2)$ using a simpler \emph{Ansatz} for the three-gluon vertex, which also reproduces the limit given in  \1eq{intgnd_u0}. Specifically, we substitute the full three-gluon vertex appearing in the second line of \1eq{Wsde_diags_Mink} by
\be 
\overline{\g}^{\,\mu}_{\mu\alpha}(-r,k,t) \to {\overline{\Gamma}_{\!0}}^{\mu}_{\;\mu\alpha}(-r,k,t)\bL(r^2,k^2)\,, \qquad \bL(r^2,k^2) = \frac{1}{2}[ \Ls(r^2) + \Ls(k^2) ] \,, 
\label{bLdef}
\ee
where $\overline{\Gamma}_{\!\!0}^{\,\alpha\mu\nu}(q,r,p)$ is the tree level equivalent of the $\overline{\g}^{\,\alpha\mu\nu}(q,r,p)$, defined in \1eq{bare3g}.
This \emph{Ansatz} amounts to substituting into  \1eq{fullK} $X_1 = X_4 = X_7 = \bL(r^2,k^2)$, with all other $X_i$ and $Y_i$ set to zero. With this approximation, $\w_2(x)$ is still given by \1eq{Wsde_diags_Euc}, but with ${\cal K}(x,y,u)$ replaced by
\be 
{\cal K}(x,y,u) \to {\cal K}_{sg}(x,y,u) = \left[ \sqrt{xy}(c_\phi^2 + 2 ) + u c_\phi \right] \bL(x,y) \,. 
\label{K_s}
\ee
Then, substituting \1eq{K_s} into \1eq{intgnd} it is straightforward to show that the limit in \1eq{intgnd_u0} is exactly reproduced\footnote{Any combination of the form $\bL(x,y)= b\Ls(x) + (1 - b)\Ls(y)$ instead of \1eq{bLdef} preserves \1eq{intgnd_u0}. The particular form used in \1eq{bLdef} has the advantage of preserving the symmetry of $\mathcal{K}(x,y,u)$ under the exchange of $x\leftrightarrow y$. We have explicitly checked that the extreme cases 
  $b=0$ and $b=1$ lead to results that are nearly identical to those obtained with $b=\frac{1}{2}$, shown in \fig{fig:intgnd_Ws}.}.

The $\w(r^2)$ that is obtained through the use of \1eq{K_s}, denoted by $\ws(r^2)$, is shown as the purple dot-dashed curve in the right panel of \fig{fig:intgnd_Ws}, where it is compared to the result obtained from \1eq{Wsde} using the general kinematics $X_i$ and $Y_i$ (blue solid line). The purple band around $\ws(r^2)$ corresponds to propagated statistical errors in the $\Ls(r^2)$
of~\cite{Aguilar:2021lke}, obtained by implementing in \1eq{bLdef} the substitution 
$\Ls(r^2)\to \Ls^\pm(r^2)$ [see \1eq{Lerr}].

In the right panel of \fig{fig:intgnd_Ws} we see that the two approximations for $\w(r^2)$ agree within the error bands, except for a small region around $3.5$~GeV. This result indicates that the error in the lattice $\Ls(r^2)$ is more important than the detailed general kinematics structure of the full three-gluon vertex, provided the limit in \1eq{intgnd_u0} is respected.

\section{\label{sec:inp} Fits for lattice inputs}

For the gluon and ghost propagators, as well as the three-gluon vertex projection $\Ls(r^2)$, we employ fits to lattice data~\cite{Bogolubsky:2009dc,Boucaud:2018xup,Aguilar:2021lke,Aguilar:2021okw}, appropriately
extrapolated to the continuum limit. The fitting functions used incorporate a number of features expected on physical grounds, particularly their asymptotic behaviors for small and large momenta. In particular: 

{\it(i)} In the UV, they reduce to the one-loop resummed behaviors dictated by renormalization-group arguments, namely
\bea 
\lim_{r^2\to\infty} \Delta^{-1}(r^2) = r^2 L_{\s{\rm{UV}}}^\delta(r^2) \,,  \;\;
\lim_{r^2\to\infty} F^{-1}(r^2) = L_{\s{\rm{UV}}}^\gamma(r^2) \,,  \;\; 
\lim_{r^2\to\infty} \Ls(r^2) =   L_{\s{\rm{UV}}}^{\delta - \gamma}(r^2)\,, 
\label{Anomalous}
\eea
where  we have defined \mbox{$ L_{\s{\rm{UV}}}(r^2) = \omega \ln\left({r^2}/{\Lambda^2} \right) $},   with  \mbox{$\omega = 11C_{\rm A}\alpha_s/(12\pi)$}.   The anomalous dimensions are given by \mbox{$\delta = 13/22$} and \mbox{$\gamma = 9/44$}.

{\it(ii)} $\Ls(r^2)$ and the derivative of the gluon propagator diverge logarithmicaly at the origin, \ie
\be 
\lim_{r^2\to 0} \Ls(r^2) = \ell \ln\left( \frac{r^2}{\mu^2} \right) \,, \qquad
 \lim_{r^2\to 0} [\Delta^{-1}(r^2)]^\prime = d \ln\left( \frac{r^2}{\mu^2} \right) \,, 
 \label{L_Delta_div}
\ee
with $\ell$ and $d$ dimensionless constants.

{\it(iii)} The $\Cfat(r^2)$ obtained from the BSE is finite at the origin~\cite{Aguilar:2011xe,Ibanez:2012zk,Aguilar:2016ock,Aguilar:2017dco}. On the other hand, the $\Ls(0)$ and $[\Delta^{-1}(0)]^\prime$ appearing in \1eq{centeuc} diverge as given by \1eq{L_Delta_div}. Moreover, it can be shown that $\w(r^2)$ has the asymptotic behavior,
\be 
\lim_{r^2\to 0}\w(r^2) = \widetilde{Z}_1 w \Delta(0) r^2 \ln\left(\frac{r^2}{\mu^2}\right) \,, 
\label{W_div}
\ee
with $w$ a dimensionless constant, such that the combination $\w(r^2)/r^2$ in \1eq{centeuc} is also logarithmicaly divergent at the origin. Consequently, consistency of \3eqs{centeuc}{L_Delta_div}{W_div} with the BSE prediction for $\Cfat(0)$ requires that all these logarithmic divergences cancel. Specifically, we demand that
\be 
\ell - F(0)\widetilde{Z}_1( w + d ) = 0 \,.
\label{divcancel}
\ee

{\it(iv)} We adopt the asymmetric MOM renormalization scheme~\cite{Aguilar:2020yni,Aguilar:2020uqw,Aguilar:2021lke,Aguilar:2020uqw},
   which imposes that 
\be 
\Delta^{-1}(\mu^2) = \mu^2 \,, \qquad F(\mu^2) = 1 \,, \qquad \Ls(\mu^2) = 1 \,, 
\label{ren_conds}
\ee
and we take the renormalization point to be $\mu = 4.3$~GeV. The fits for the lattice ingredients are all required to reduce exactly to the above values at $\mu$.
In order to incorporate all the above features, the fitting functions have rather elaborate forms.

Starting with $F(r^2)$, an accurate fit to the lattice data is obtained with
\be 
F^{-1}(r^2) = A^\gamma(r^2) + R(r^2) \,, 
\label{ghost_fit}
\ee
where $A(r^2)$,
\be 
A(r^2) := 1 + \omega \ln\left(  \frac{r^2 + \eta^2(r^2) }{\mu^2 + \eta^2(r^2)} \right) \,, 
\label{UVlog}
\ee
with
\be
\label{eta}
\eta^2(r^2) = \frac{\eta_1^2}{1 + r^2/\eta_2^2}\,,
\ee
while $R(r^2)$ is a combination of   rational functions,
\be
\label{rational} 
R(r^2) =\frac{ b_{0} + b_{1}^2 r^2 }{1 + (r^2/ b_{2}^2 )  +  (r^2/b_{3}^2)^2 } - \frac{ b_{0} + b_{1}^2 \mu^2 }{1 + (\mu^2/ b_{2}^2 )  +  (r^2/b_{3}^2)^2 } \,.
\ee
Note that $R(r^2)$ vanishes quickly at infinity, and that $R(\mu^2) = 0$ and $A(\mu^2) = 1$, enforcing the renormalization condition in \1eq{ren_conds}. Moreover, while $A(r^2)$ saturates to a constant at the origin, in the UV it recovers the perturbative logarithm, since $\eta^2(r^2)\to 0$ at large $r^2$, such that 
\be 
\lim_{r^2\to \infty} A(r^2) = L_{\s{\rm{UV}}}(r^2)\,,
\ee
where the function  $L_{\s{\rm{UV}}}(r^2)$ was defined in Eq.~(\ref{Anomalous}) with  $\Lambda^2 = \mu^2e^{-1/\omega}$.

Turning to $\Delta(r^2)$, a form that satisfies all the required conditions is
\be 
\Delta^{-1}(r^2) = r^2 \left[ \frac{d}{ 1 + ( r^2/\kappa^2) }\ln\left( \frac{r^2}{\mu^2} \right) + A^\delta(r^2) \right] + \nu^2 R(r^2) \,, \label{gluon_fit}
\ee
where the unprotected logarithm in the first term in brackets describes the IR divergence of $[\Delta^{-1}(0)]^\prime$ and drops out in the UV, while $A(r^2)$ and $R(r^2)$ are given by \2eqs{UVlog}{rational}. The parameter \mbox{$\nu = 1$~GeV} serves only to make the dimensionality of $R$ consistent with that of $\Delta^{-1}(r^2)$, without changing the dimensions of the $\sigma_i$ parameters in \1eq{rational}.

As for for $\Ls(r^2)$ we use the fitting form
\be 
\Ls(r^2) =  \frac{\ell}{ 1 + ( r^2/\kappa^2) }\ln\left( \frac{r^2}{\mu^2} \right) + A^{\delta - \gamma}(r^2) + R(r^2) \,, 
\label{L_fit}
\ee
with $A(r^2)$ and $R(r^2)$ given by \2eqs{UVlog}{rational}, respectively.

Note that although in \mbox{Eqs. ~\eqref{ghost_fit},~\eqref{gluon_fit}, ~and~\eqref{L_fit}} we use the same names for the parameters 
$\kappa^2$, $\eta^2_i$,  $b_0$,  and  $b^2_i$,  for economy,  they are allowed to assume different values for each of the functions $F(r^2)$, $\Delta(r^2)$,  and $\Ls(r^2)$.

Next, the coefficient $\ell$ in \1eq{L_Delta_div}, characterizing the rate of divergence of $\Ls(0)$, has been determined from lattice results to be \mbox{$\ell = 0.11$}~\cite{Aguilar:2021lke}, and is held fixed during the fitting procedure. In contrast, the rate of divergence $d$ of $[\Delta^{-1}(r^2)]^\prime$ is \emph{not} accurately determined from the lattice, since the derivative is sensitive to the larger lattice noise in $\Delta(r^2)$ in the deep IR. Moreover, the coefficient $w$ in \1eq{W_div} depends on the ingredients, including $\Delta(r^2)$, used in the SDE evaluation of $\w(r^2)$ through \2eqs{Wsde}{Wsde_diags_Euc}. As such, in order to enforce \1eq{divcancel}, $w$ and $d$ have to be varied simultaneously, until the cancellation of the divergences has been reached to acceptable precision. 

\begin{figure}[t]
 	\includegraphics[width=0.31\textwidth]{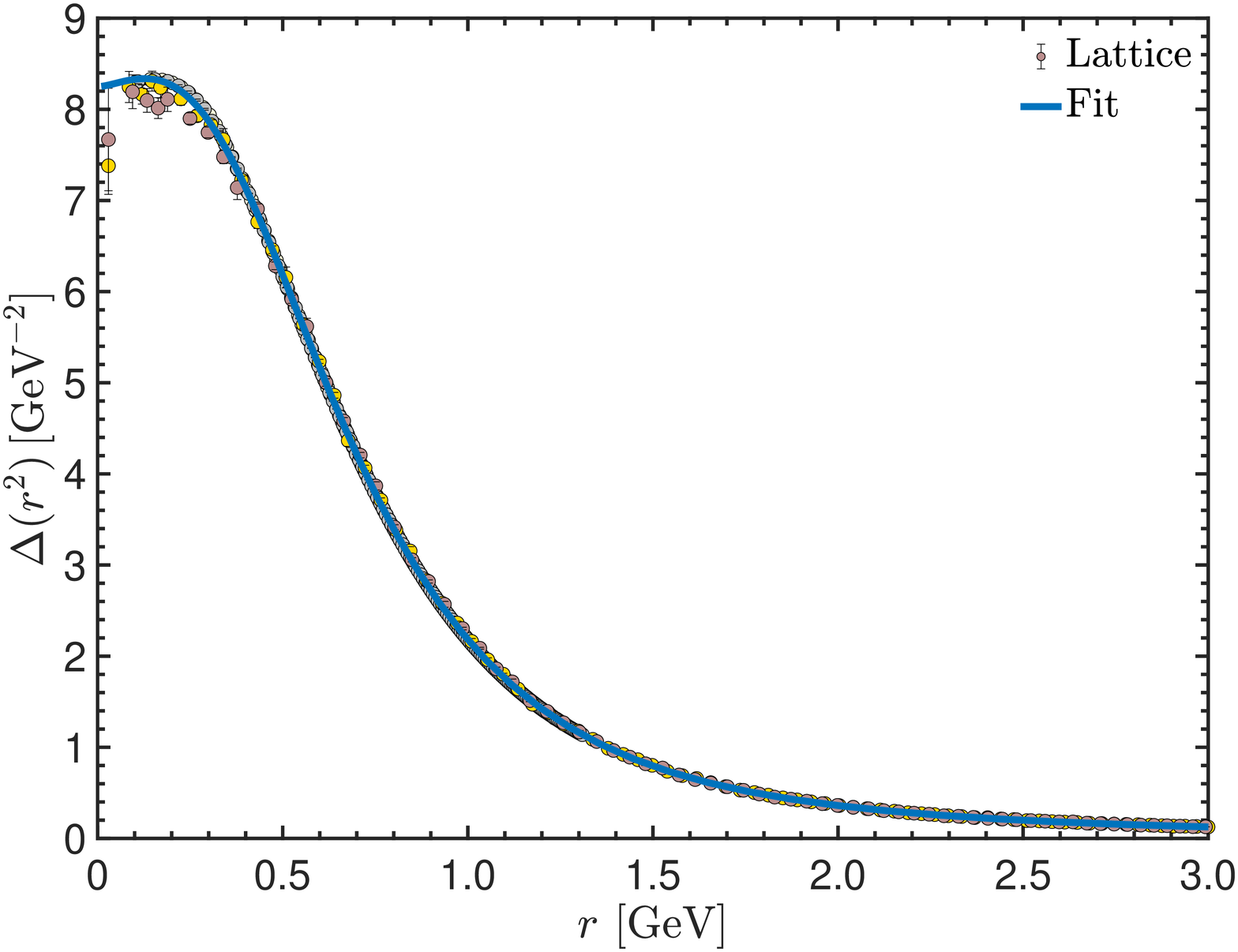}\hfill
	\includegraphics[width=0.31\textwidth]{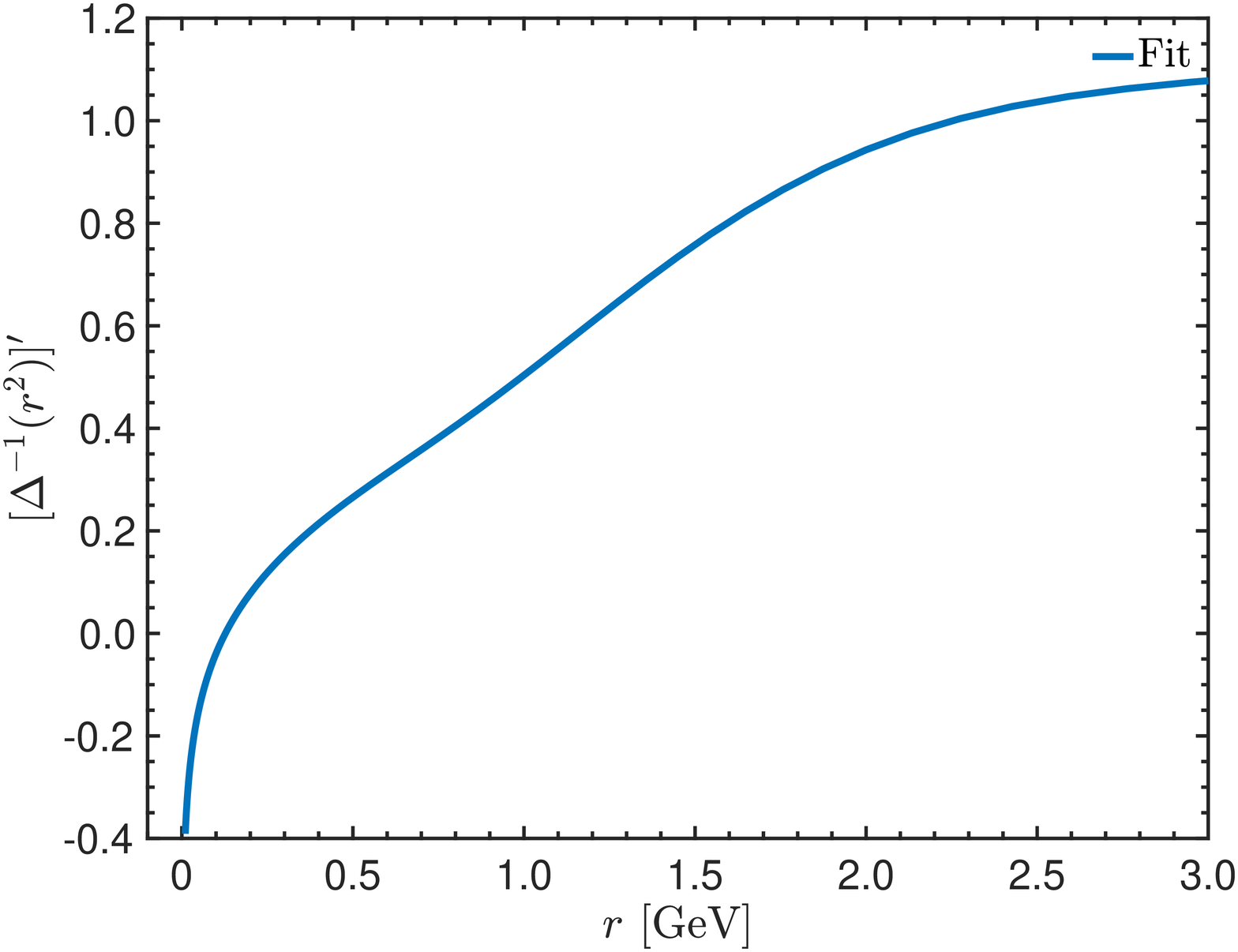}\hfill
 	\includegraphics[width=0.31\textwidth]{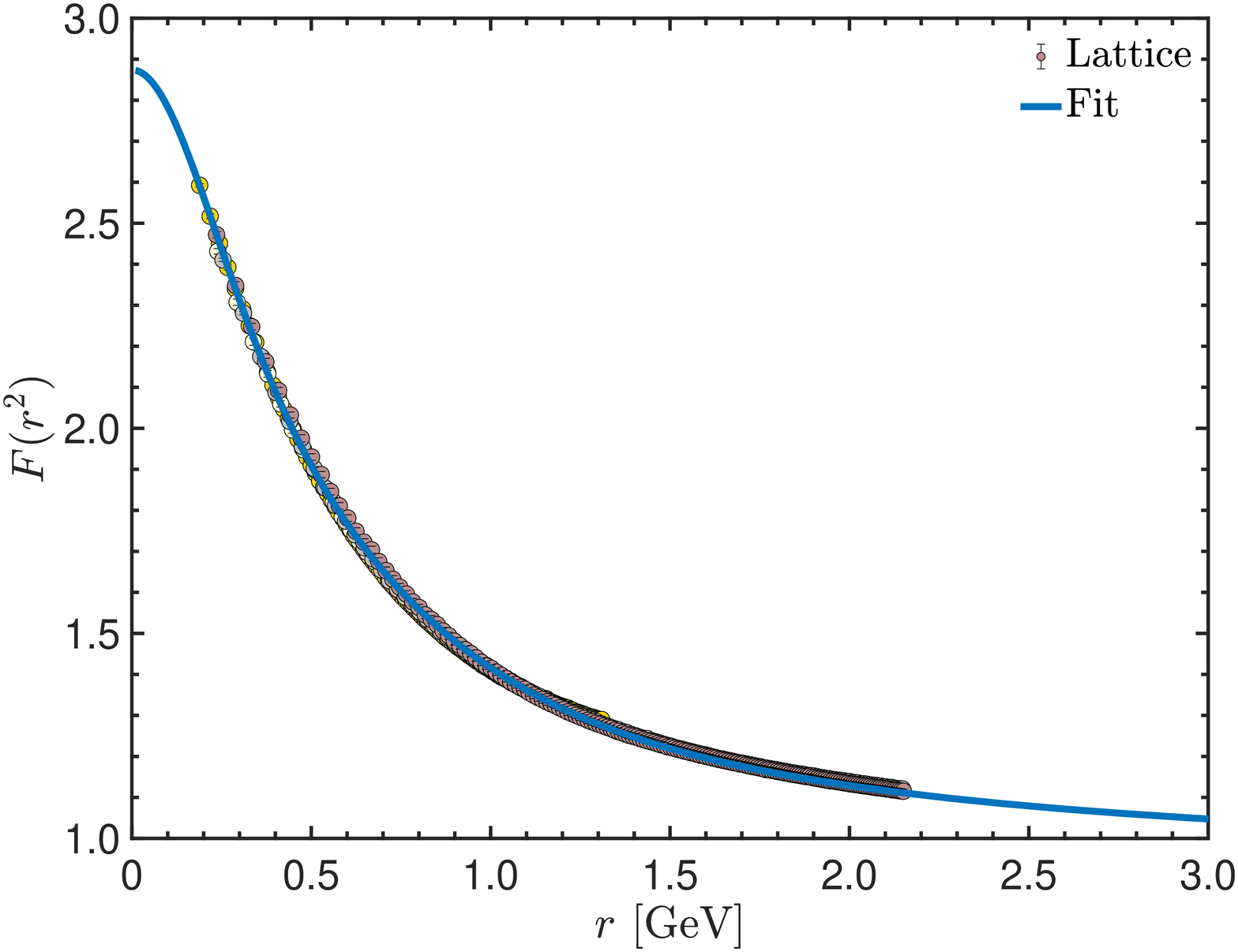} 		
 	\caption{ Left: Lattice data of \cite{Bogolubsky:2009dc,Aguilar:2021okw} (points), and the fit of \1eq{gluon_fit} (blue solid line), for the gluon propagator, $\Delta(r^2)$. Center: The derivative $[\Delta^{-1}(r^2)]^\prime$, obtained through differentiation of \1eq{gluon_fit}. Right: Lattice data (points) of \cite{Boucaud:2018xup}, and the fit of \1eq{ghost_fit} (blue solid), for the ghost dressing function, $F(r^2)$. }
\label{fig:fits}
\end{figure}

The fitting parameters resulting for $\Delta(r^2)$, $F(r^2)$, and $\Ls(r^2)$ are given in Table~\ref{pars} and its caption. The resulting curves for $\Delta(r^2)$ and $F(r^2)$ are compared to the lattice data of \mbox{\cite{Bogolubsky:2009dc,Aguilar:2021okw}}  in \fig{fig:fits}, where we also show \mbox{$[\Delta^{-1}(r^2)]^\prime$}. The lattice data of \cite{Aguilar:2021lke} and corresponding fit for $\Ls(r^2)$ are shown as the points and black continuous curve, respectively, in the right panel of \fig{fig:Cres}. 

Comparing the curve of $\Ls(r^2)$ in \fig{fig:Cres} to that of \mbox{$[\Delta^{-1}(r^2)]^\prime$} in \fig{fig:fits}, we see that $[\Delta^{-1}(r^2)]^\prime$ is responsible for reproducing the overall shape of $\Ls(r^2)$ in the WI of \1eq{centeuc}, with the other ingredients providing minor quantitative modulations.

Now, it is clear from Figs.~\ref{fig:Cres} and \ref{fig:fits} that the lattice quantity with the largest error in the present analysis is 
$\Ls(r^2)$.  In order to propagate the error of $\Ls(r^2)$ to other quantities that depend on it, we make a band around $\Ls(r^2)$, delimited by
\be 
\Ls^\pm(r^2) = \Ls(r^2) \pm \frac{\delta_1}{ 1 + ( r^2/\delta_2^2 )^2 } \,, 
\label{Lerr}
\ee
with parameters \mbox{$\delta_1 = 0.08$} and \mbox{$\delta_2^2 = 5$~GeV$^2$}.

\begin{table}
\begin{tabular}{|c|c|c|c|c|c|c|c|}
\hline
 & $\kappa^2$ [GeV$^2$] & $\eta_1^2$ [GeV$^2$] & $\eta_2^2$ [GeV$^2$] & \hphantom{       } $b_0$ \hphantom{       } & $b_1^2$ [GeV$^{-2}$] & $b_2^2$ [GeV$^2$] &  $b_3^2$ [GeV$^2$] \\
\hline
$F(r^2)$ & - & 3.60 & 0.148 & -0.566 & 0.004 & 0.375 & 24.2 \\
\hline
$\Delta(r^2)$ & 1.33 & 0.889 & 2.570 & 1.254 & 0.723 & 1.553 & 2.08 \\
\hline
$\Ls(r^2)$ & 18.2 & 0.200 & 6.36 & 0.241 & -0.646 & 0.310 & 1.28 \\
\hline
\end{tabular}
\caption{ Fitting parameters $\kappa^2$, $\eta^2_i$,   $b_0$, and $b^2_i$, to be used in each of Eqs. ~\eqref{ghost_fit}, ~\eqref{gluon_fit},  and \eqref{L_fit} for $F(r^2)$, $\Delta(r^2)$,  and $\Ls(r^2)$, respectively.  In addition to the values below, we set \mbox{$d = 0.075$} and 
\mbox{$\ell = 0.11$},  in \2eqs{gluon_fit}{L_fit}, respectively,  and \mbox{$\omega = 0.214$} in the Eqs.~\eqref{ghost_fit}, \eqref{gluon_fit},  and \eqref{L_fit}. }
\label{pars} 
\end{table}

Lastly, for the value of $\alpha_s$ in the asymmetric MOM scheme, which appears in the SDE for $\w(r^2)$ given by \1eq{Wsde}, we use the value \mbox{$\alpha_s = 0.27$} determined by lattice simulations~\mbox{\cite{Athenodorou:2016oyh,Boucaud:2017obn}}. For the $F(0)$ appearing in \1eq{centeuc} we obtain from \1eq{ghost_fit} and Table~\ref{pars} the value \mbox{$F(0) = 2.88$}.

%


\end{document}